%% file: Paper.tex
\documentclass[a4paper,fleqn,usenatbib]{mnras}

\usepackage[T1]{fontenc}
\usepackage{ae,aecompl}

\usepackage{graphicx}	
\usepackage{url,hyperref}
\usepackage{txfonts}
\usepackage{footmisc}

\input{commands.tex}

\newcommand{\graynoun}{$\gamma$ ray}
\newcommand{\gray}{$\gamma$-ray}

\title[VHE $\gamma$-ray emission from two blazars]{Very high energy $\gamma$-ray emission from two blazars of unknown redshift and upper limits on their distance} 

\clearpage
\input{authors.tex}
\clearpage

\date{Accepted XXX. Received YYY; in original form ZZZ}

\pubyear{2020}

\begin{document}
\label{firstpage}
\pagerange{\pageref{firstpage}--\pageref{lastpage}}
\maketitle

\clearpage
\input{affiliations.tex}
\clearpage

\begin{abstract}
We report on the detection of very-high-energy (VHE; $E > 100$~GeV) \gray\ emission from the BL Lac objects \kuv\ and \pks\ with the High Energy Stereoscopic System (\hess). \hess\ observations were accompanied or preceded by multi-wavelength observations with \fermi, XRT and UVOT on board the \swift\ satellite, and ATOM. Based on an extrapolation of the \fermi\ spectrum towards the VHE \gray\ regime, we deduce a 95\% confidence level upper limit on the unknown redshift of \kuv\ of $z < \KuvUppz$, and of  \pks\ of $z < \PKSUppz$. When combined with previous spectroscopy results the redshift of \kuv\ is constrained to $\KuvLowz \leq z < \KuvUppz$ and for \pks\ to $\PKSLowz \lessapprox z < \PKSUppz$. 
\end{abstract}




\section{Introduction}

The extragalactic \gray\ sky is strongly dominated by blazars, a class of radio-loud, jet-dominated active galactic nuclei (AGN) in which the relativistic jet is oriented at a small angle with respect to the line of sight.
This alignment leads to strong Doppler enhancement of the observed flux across the entire electromagnetic spectrum as well as a shortening of the apparent variability, which has been observed to be as short as a few minutes \citep[e.g.,][]{1996Natur.383..319G,Albert07,Aharonian07,Arlen13}.
The radio to \graynoun s spectral energy distribution (SED) of blazars is dominated by two distinct, non-thermal radiation components, where the low-energy emission (from radio through UV or X-rays) is generally ascribed to synchrotron radiation from relativistic electrons/positrons in the jet.
For the high-energy emission, both a Compton-scattering-based leptonic scenario and a hadronic scenario, where \gray\ emission results from proton synchrotron radiation and photo-pion-induced processes, are plausible \citep[see, e.g.,][for a comparative study of both types of models]{Boettcher13}. 

The multi-wavelength variability of blazars exhibits complex patterns on all time scales, from years down to minutes, with variability at different frequencies sometimes being correlated, but sometimes also showing uncorrelated behaviour, such as ``orphan'' \gray\ flares without significant counterparts at lower frequencies \citep[e.g.,][]{Krawczynski04}. 
The fast (intra-day) variability time scales indicate that the broadband emission, at least from optical to \gray\ frequencies, must originate in small, localised regions along the jet, with a radius no larger than ${c\, \tau\, \delta\, (1+z)^{-1} \simeq 1.8\times10^{14}\, \tau_{\rm{10 min}}\, \delta_{10} (1+z)^{-1}}$~cm, where $\tau_{\rm{10 min}} = \tau/\rm{10 \; min}$ is the observed variability time-scale, ${\delta_{10} = \delta / 10}$ is the Doppler factor of the emitting region, and $z$ is the redshift of the source.
Measurements of blazar SEDs and their variability thus provide a unique laboratory for probing the microphysical processes of particle acceleration and radiative cooling in AGN jets. 

Based on the location of the peak of the low-energy (synchrotron) spectral component, $\nu_{\rm sy}$, blazars are sub-divided into low-synchrotron-peaked (LSP, with $\nu_{\rm sy} \le 10^{14}$~Hz), intermediate-synchrotron-peaked (ISP, with $10^{14} \, {\rm Hz} < \nu_{\rm sy} \le 10^{15}$~Hz), and high-synchrotron-peaked (HSP) blazars \citep[with $\nu_{\rm sy} > 10^{15}$~Hz;][]{Abdo10}.
Blazars are also divided into BL Lacertae objects (BL Lacs) and Flat-Spectrum Radio-Quasars (FSRQs) on the basis of their optical/UV spectrum, which is almost featureless in BL Lacs, and shows broad emission lines in FSRQs.
While FSRQs are all LSPs, BL Lacs are characterised by a variety of peak frequencies. 
The vast majority of blazars detected by ground-based Imaging Atmospheric Cherenkov telescope (IACT) facilities in very high energy (VHE; $E \ge 100$~GeV) \graynoun s are BL Lacs belonging to the HSP class.

As VHE \graynoun s are subject to $\gamma\gamma$ absorption due to $e^+e^-$ pair production on IR -- optical photons, the extragalactic background light (EBL) limits the cosmic horizon out to which VHE \gray\ sources are detectable \citep[e.g.,][]{Nikishov62,GS67,Stecker92,Aharonian06,Finke10}. 
In agreement with this expectation, no VHE \gray\ source has so far been detected at a redshift $z > 1$, the current record holder being the gravitationally-lensed blazar S3~0218+35 at a redshift of $z = 0.944$ \citep{Ahnen16}.

While the EBL absorption represents an intrinsic limit for VHE astrophysics, it is possible to probe the EBL itself through its absorption imprint on VHE spectra of blazars \citep[e.g.,][]{HESSEBL, FermiEBL, VERITASEBL, MAGICEBL}. 

For such studies, as large a sample of VHE blazars as possible, over as large a redshift range as possible, is necessary in order to disentangle source-intrinsic high-energy cut-offs from the effect of absorption on the EBL, improve the statistical uncertainty on the measurement, and study the evolution of the EBL with the redshift. 
A high redshift by itself does not necessarily mean that a blazar is interesting for propagation studies, given that both the intrinsic brightness of the source and the hardness of its spectrum play a key role.
This motivates continued programmes by all currently operating IACT arrays to detect new VHE blazars and characterise their spectral properties in the VHE band.
The selection of targets for such searches is most commonly based on an extrapolation of the High Energy (HE; $E > 100$~MeV) \gray\ spectra as measured by the {\it Fermi} Large Area Telescope \citep[LAT,][]{2fgl,3fgl}, taking into account the expected attenuation due to EBL $\gamma\gamma$ absorption: bright and hard LAT sources are prime candidates for pointed observations with IACTs.

A major problem when studying blazars is that the measurement of a BL Lac's redshift is not trivial, due to the weakness of the emission lines (if any) in the optical spectrum. 
It is possible to constrain the redshift via the identification of absorption lines from the host-galaxy, or to provide a lower limit on it via the detection of absorption lines from inter-galactic absorbers. 
If the redshift is unknown, it is possible to use the current knowledge of the EBL together with the information from \fermi\ to compute an upper limit on the redshift of the source.

In this paper we present the detection with \hess\ in the VHE regime of two blazars, selected for their hard \fermi\ spectra and large upper limits on their redshift, \kuv\ and \pks. We present new constraining upper limits on their distance using \fermi\ and \hess\ spectral information. 
We describe the details of the \hess\ observations and data analysis in Section~\ref{HESS}, and results from multi-wavelength observations in Section~\ref{multi}. 
Then in Section~\ref{sed} we present the SEDs, and in Section~\ref{redshift} we discuss our redshift constraint. 
We conclude with a summary of our results in Section~\ref{summary}. 
Throughout this text, the results for \kuv\ and \pks\ are presented in separate subsections.

\subsection{KUV 00311-1938} \label{KUVintro}

\kuv\ has been classified as a BL Lac firstly by a spectroscopic identification in the sample of bright, soft, high-Galactic-latitude X-ray sources from the ROSAT All-Sky Survey \citep{1999yCat.9010....0V, 1998A&A...335..467T}.
Later, \cite{2000ApJS..129..547B} associated this bright ($S_{\rm X} = 1.32^{-11} \rm ergs \; s^{-1} \; cm^{-2}$) ROSAT source with a strong radio emitter \citep[][$S_{1.4 \; \rm GHz} = \rm 18.8 \; mJy$]{1998AJ....115.1693C}.
Its extreme value of the X-ray to radio flux ratio and its high X-ray flux, led to its inclusion in the Sedentary Multi-Frequency Survey catalogue \citep{2007A&A...470..787P}, which primarily selected HSPs. 

A first evaluation of the redshift of \kuv\ was performed by \cite{2007A&A...470..787P}, where a value of 0.61 was quoted although flagged as ``tentative''.
Later, \cite{2014A&A...565A..12P} detected the MgII doubled with the X-Shooter spectrograph operating on the VLT to estimate a secure lower limit on the redshift of the source to 0.506, and used non-detection of the host galaxy to place an upper redshift limit of 1.54.
Recently, \cite{ICRC2019} attempted to constrain the redshift of \kuv, by estimating the range of EBL absorption allowed by the preliminary \hess\ observations presented in \citet{2012AIPC.1505..490B} and matching this to existing EBL models. 
They conclude that a redshift around 0.5-0.6 is the most plausible.

\kuv\ has been imaged multiple times with VLBI by \cite{2018ApJ...853...68P}, who observed superluminal apparent motions of a secondary jet component with $\beta_{\rm app} = 6 \pm 2$ relative to a fixed radio core at \RAVLA, \DECVLA \citep{2014ApJ...797...25P}. 
We take the position of the radio core as our nominal location for \kuv.

The \fermi\ \citep{2009ApJ...697.1071A} reported the detection of a very bright HE \gray\ source consistent with \kuv\ in all the catalogues, including the ones compiled with high energy events only \citep{2015ApJS..218...23A, 2017ApJS..232...18A}.

In the most-recent 4FGL \citep{2019arXiv190210045T} catalog this source has an integrated flux of \FermiFourFGLFlux\ in the [1--100] GeV range and a photon power-law spectrum with a hard (\FermiFourFGLPLIndex) spectral index. 
The spectrum is reported as curved, with both a log-parabola (\FermiFourFGLLPSig) as well as a power-law with exponential cutoff (\FermiFourFGLECSig) being preferred over the simple power-law. 
The 4FGL reports modest fractional variance on both yearly ($F_{var} = $ \FermiFourYearVar) and bi-monthly ($F_{var} = $ \FermiFourTwoMonthVar) timescales, with the larger variability seen on the shorter timescale. 

\hess\ observations of this source started at the end of 2009 and were pursued until 2014, leading to the detection of VHE \g-ray emission from this distant BL Lac (see Sec.\ \ref{HESS}), reported here.

\subsection{PKS 1440-389}  \label{PKSintro}

\pks\ \citep[\PKSRAnom\, \PKSDECnom\ ;][]{Jackson02} was first detected as a bright radio source in the Parkes survey \citep{WO90}. 
The source has been observed repeatedly by the TANAMI \citep[Tracking of AGN with Austral Milliarcsecond Interferometry,][]{Ojha10} project between 2010 and 2016. 
\cite{Krauss16} present three quasi-simultaneous SEDs of PKS~1440-389, which indicate a synchrotron peak at $10^{15}$~Hz$\, \lesssim \nu_{\rm sy} \lesssim 10^{16}$~Hz, consistent with an HSP classification. 

In the first data release of the 6dF Galaxy Survey \citep{2004MNRAS.355..747J}, the redshift of the source is listed as $z=0.065$, but this redshift value is no longer included in the final version of the 6dF catalogue \citep{2009MNRAS.399..683J} due to the poor quality of the optical spectrum. 
Despite many follow-up observations in different wavelength regimes, the redshift of \pks\ remains uncertain due to its featureless continuum spectrum \citep[e.g.,][]{2015AJ....149..163L}.

The current constraint from optical spectroscopic observations is $0.14 < z < 2.2$ \citep{2013ApJ...764..135S}.

\pks\ stands out as a bright \gray\ HSP with a hard, well-constrained \fermi\ spectrum, with the 4FGL reporting a power-law index of \FermiPKSFourFGLPLIndex\ and an integrated flux (over [1--100] GeV) of \FermiPKSFourFGLFlux. 
Similar to \kuv\, here too the spectrum shows evidence of being curved, with both a log-parabola (\FermiPKSFourFGLLPSig) as well as a power-law with exponential cutoff (\FermiPKSFourFGLECSig) being preferred over the simple power-law. 
The 4FGL also reports modest variability on both yearly ($F_{var} = $\FermiPKSFourYearVar) and bi-monthly ($F_{var} = $ \FermiPKSFourTwoMonthVar) timescales.

Assuming a redshift near the lower limit of the allowed range, the EBL-corrected extrapolation of the \fermi\ spectrum into the VHE regime appeared promising for detection, and \hess\ observations in 2012 yielded the discovery of the source at VHE \citep{2015ICRC...34..862P}.
Using that preliminary \hess\ spectrum, \cite{2019ApJ...884L..17S} computed a model-dependent limit on the distance of \pks, $ 0.14 < z < 0.24 $, assuming a hadronic origin of the emission.


\section{\label{HESS}H.E.S.S. data analysis and results}

\hess\ is an array of five IACTs located in the Khomas Highland in Namibia (S~$23^\circ 16' 18''$, E~$16^\circ 30' 00''$) at an altitude of about 1800~m above sea level. 
From January 2004 to October 2012, the array was a four-telescope instrument, with telescopes CT1-4.
Each of the telescopes, located at the corners of a square with a side length of 120 m, has a mirror surface area of 107 $\rm m^{2}$ and a camera composed of $960$ photomultipliers covering a large field of view (FoV) of $5^{\circ}$ diameter.
The stereoscopic system works in a coincidence mode, requiring at least two of the four telescopes to trigger the detection of an extended air shower.
In its initial four-telescope configuration used here, \hess\ is sensitive to \gray\ energies from 100~GeV to about 100~TeV \citep{Aharonian06Crab}.

In October 2012, a fifth telescope (CT5), with a mirror surface area of 600 $\rm m^{2}$ and an improved camera \citep{2014NIMPA.761...46B} was installed at the centre of the original square.

All observations were done in wobble mode, where the source is observed with an offset of $0.5^{\circ}$ with respect to the centre of the instrument's field of view to allow for simultaneous background measurements \citep{Fomin94}.
The analysis of the \g-ray emission from the two sources was carried out with the analysis procedure described in \cite{2014APh....56...26P}, where an enhanced low-energy sensitivity with respect to standard analysis methods \citep{Aharonian06Crab} is achieved.
Since these sources are potentially very distant and so likely have very soft spectra, a special analysis configuration with a charge value of $\rm 40$ photo electrons is used as the minimal required total amplitude for the cleaned and reconstructed image in each telescope.

The statistical significance of the two detections was determined using the Reflected background modelling method \citep{Aharonian06Crab} and Eq. (17) of \citet{LiMa}. In the Reflected background method, $\alpha$ is the reciprocal of the number of OFF-source regions considered in a run. If the number of OFF regions vary from run to run, the average $\alpha$ for all the runs is used.

The time-averaged differential VHE \g-ray spectra of the sources was derived using the forward-folding technique described in \cite{Piron}. The maximum energy for both fits was chosen to be 3 TeV, while the minimum energy was left free and therefore represents the threshold energy.

The systematic uncertainties have been estimated following the procedure described in \cite{Aharonian06Crab}, with the uncertainty from the selection cuts estimated using the difference between the lead and cross-check analysis. A fit of a log parabola did not significantly improve the fit.

\begin{table}
  \centering
  \caption{Results of the spectral fitting of the \hess\ data for the blazars \kuv\ and \pks. The threshold energy, $E_{\rm th}$, is defined as the energy at which the energy bias is less than 10 \%. 
  $E_0$ is set to be the decorrelation energy of the fit. 
  We give the integrated fluxes for both sources above the larger value of $E_{\rm th}$.}
  \label{tab:hessfit}
  \begin{tabular}{ll}
	\hline
	\hline
	\kuv\ & \\
	\hline
	\hline
	$\Gamma$    &    $\KuvIndexValue$ \\ 
   $N_0$       &    $\KuvFluxAtDecValue  \tevcms$ \\
  $E_0$ &     \KuvEdec  \\
  $E_{\rm th}$ &     \KuvEth  \\
  $F_{ (E>E_{\rm th}) }$ & \KuvFluxOverThreshValue \cms \\
  $F_{(E > \PksEth)}$  & \KuvCrabatPKS\ Crab flux $> \PksEth$ \\
\hline
\\
	\hline
	\hline
	\pks\ & \\
	\hline
	\hline
	$\Gamma$    &    $\PksIndexValue$ \\ 
   $N_0$       &    $\PksFluxAtDecValue  \tevcms$ \\
  $E_0$ &     \PksEdec  \\
  $E_{\rm th}$ &     \PksEth  \\
  $F_{(E>E_{\rm th})}$ & \PksFluxOverThreshValue \cms \\
  $F_{(E > \PksEth)}$ &  \PKSHessCrabFlux\ Crab flux $> \PksEth$ \\

\hline
  \end{tabular}
\end{table}

\subsection{\kuv}

\hess\ started observing several high-redshift blazars in the last years of its 4-telescope configuration and continued after the addition of a fifth telescope.
Among the blazars observed by \hess, \kuv\ had the largest lower limit on $z$, and preliminary results of the observations were published in \cite{2012AIPC.1505..490B}.

Observations of \kuv\ were carried out with \hess\ in a campaign between end of 2009 and end of 2014 (MJD 55145--56954), leading to \KuvTime\ hours of good-quality data \citep[after hardware and weather quality selection criteria were applied with a procedure similar to that described in][]{Aharonian06Crab}. These observations were taken with an average zenith angle of \HessMeanZenit.

To keep consistency in the analysis of later data and the data taken mostly before the fifth telescope was added to the array, all data are analysed in the four telescope configuration, removing data from CT5.

The source is detected at a level of \KuvSig\ standard deviations, with an excess of \KuvExcess\ counts from the ON region of $0.1^{\circ}$ radius centred at the nominal position of the source. The total number of ON- and OFF-source events are $\rm N_{ON} = \KuvOn$ and $\rm N_{OFF} = \KuvOff$, with a background normalisation factor $\alpha = \KuvAlpha$.
A fit to the uncorrelated excess map yields a position for the excess of \KUVRAfitted\ and \KUVDECfitted, consistent with the position of the radio core seen by the VLA.
The systematic uncertainty is estimated as \HESSRAsys\ and \HESSDECsys, smaller than the statistical one.

The light curve for \kuv, assuming a fixed index and binned by observation period (between August and December), is shown in Fig. \ref{fig:LightCurve}.
A fit of a constant to these flux points finds no significant deviation from a steady flux ($\chi^{2}/{\rm ndf} = 4.2/5$), nor is any variation found at daily timescales.

The \kuv\ time-averaged spectrum is presented in Fig.\ \ref{fig:kuvspectrum}, and the fit results are presented in \autoref{tab:hessfit}. 
The spectrum was fitted by a  power-law function, see \autoref{tab:latshapes} for the full expression. 

All the results have been cross-checked and confirmed with the analysis method in \citet{DNR09}.

\subsection{\pks}
\hess\ observations of \pks\ were conducted during the 3-month period between February 28 and May 27, 2012 (MJD~$55985-56074$) at a mean zenith angle of $17^{\circ}$.
After quality selection and dead time correction, the data sum up to a total observation time of 11.9~hours. 

The source is detected at a level of \PksSig\ standard deviations, with an excess of 342 counts from the ON region of $0.1^{\circ}$ radius centred at the nominal position of the source. The total number of ON- and OFF-source events being $\rm N_{ON} = 999$ and $\rm N_{OFF} = 6391$, respectively, with a background normalisation factor $\alpha = 0.102754 $. 

A fit to the uncorrelated excess map yields a position for the excess of \PKSRAfitted\ and \PKSDECfitted, which is spatially consistent with the radio position of the BL Lac object \pks\ \citep{Jackson02}. 

\begin{figure*}
  \hspace{-0.4cm}
  \includegraphics[width=1.1\columnwidth]{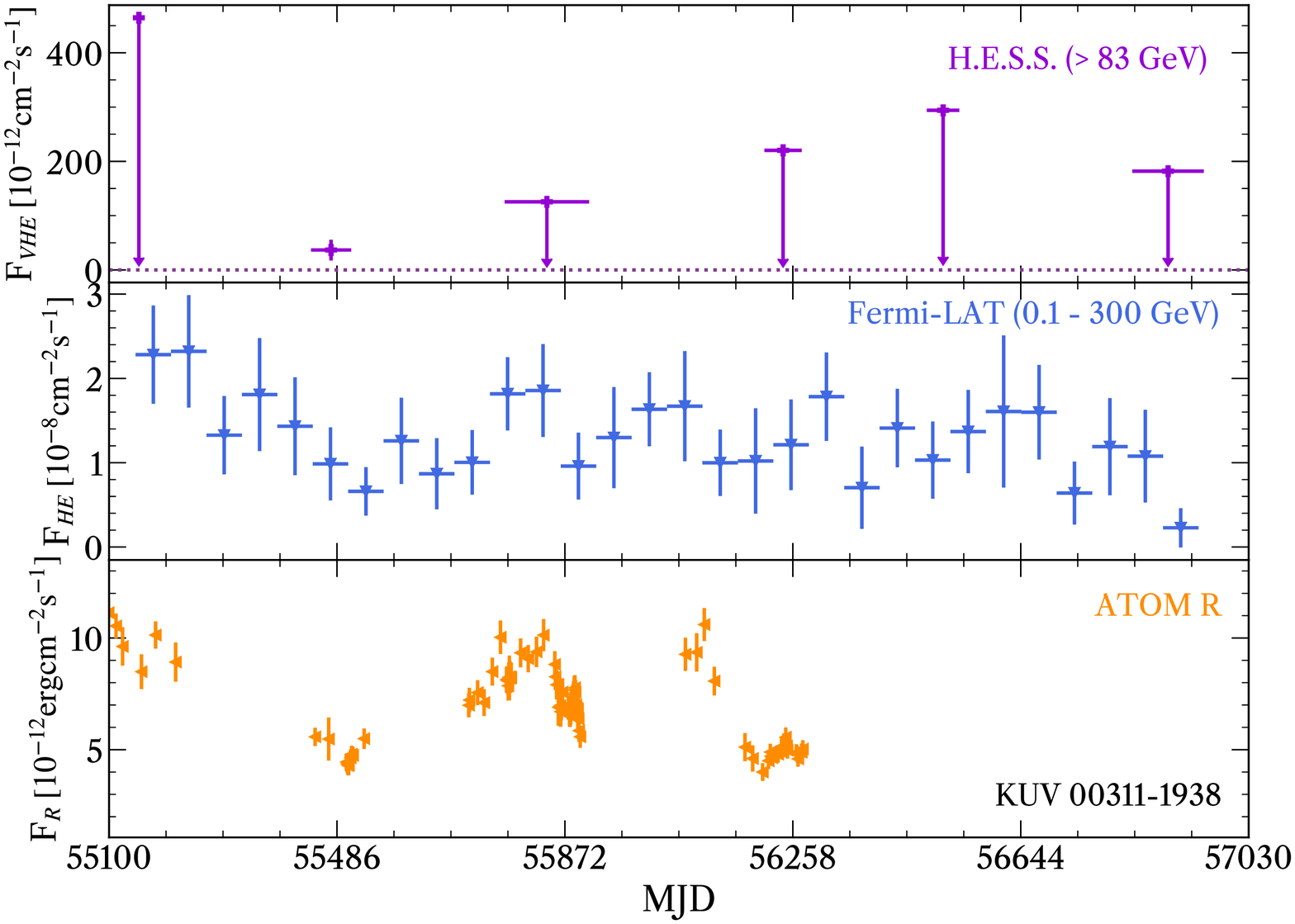}
  \hspace{-0.9cm}
  \includegraphics[width=1.1\columnwidth]{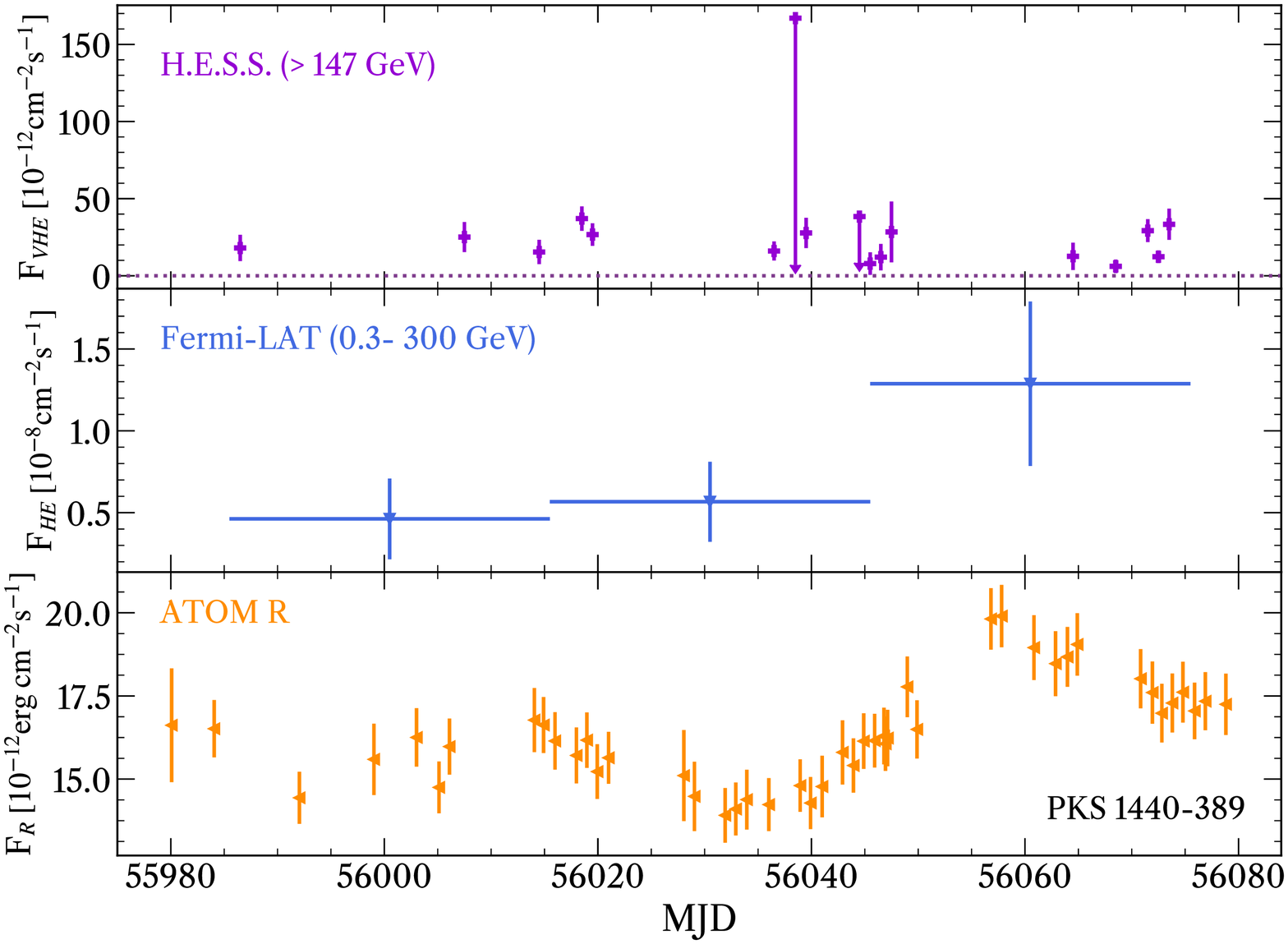}
  \caption{
  Light curves for \kuv\ (left) and \pks\ (right). 
  \textsc{Top panels.} \hess\ light curves showing the flux above $\rm 0.83 \, TeV$ per observing season (left) and above $\rm 0.147 \, TeV$ with daily binning (right). 
  Only statistical errors are shown, and upper limits are calculated when the significance in a bin is below 1$\sigma$.
  \textsc{Middle panels.} Light curve of \fermi\ observations in the energy range \KuvFermiEnergyRange\ in 2-month binning (left). 
  \fermi\ light curve in the energy range \PksFermiEnergyRange\ in monthly bins (right). The significance in each bin is at least 3$\sigma$.
  \textsc{Bottom panel.} Light curve of \atom\ observations with the R filter.
  }
  \label{fig:LightCurve}
\end{figure*}

A daily binned light curve was derived for energies above the energy threshold (147~GeV), assuming a fixed spectral index, and is shown in the top right panel of \reffig{fig:LightCurve}.
A fit with a constant to the flux points showed no significant deviation from a steady flux ($\chi^{2}/{\rm ndf} = 19.41/14$), nor was any variation detected on monthly time scales.

\begin{figure*}
  \hspace{-.4cm}
  \includegraphics[width=1.11\columnwidth]{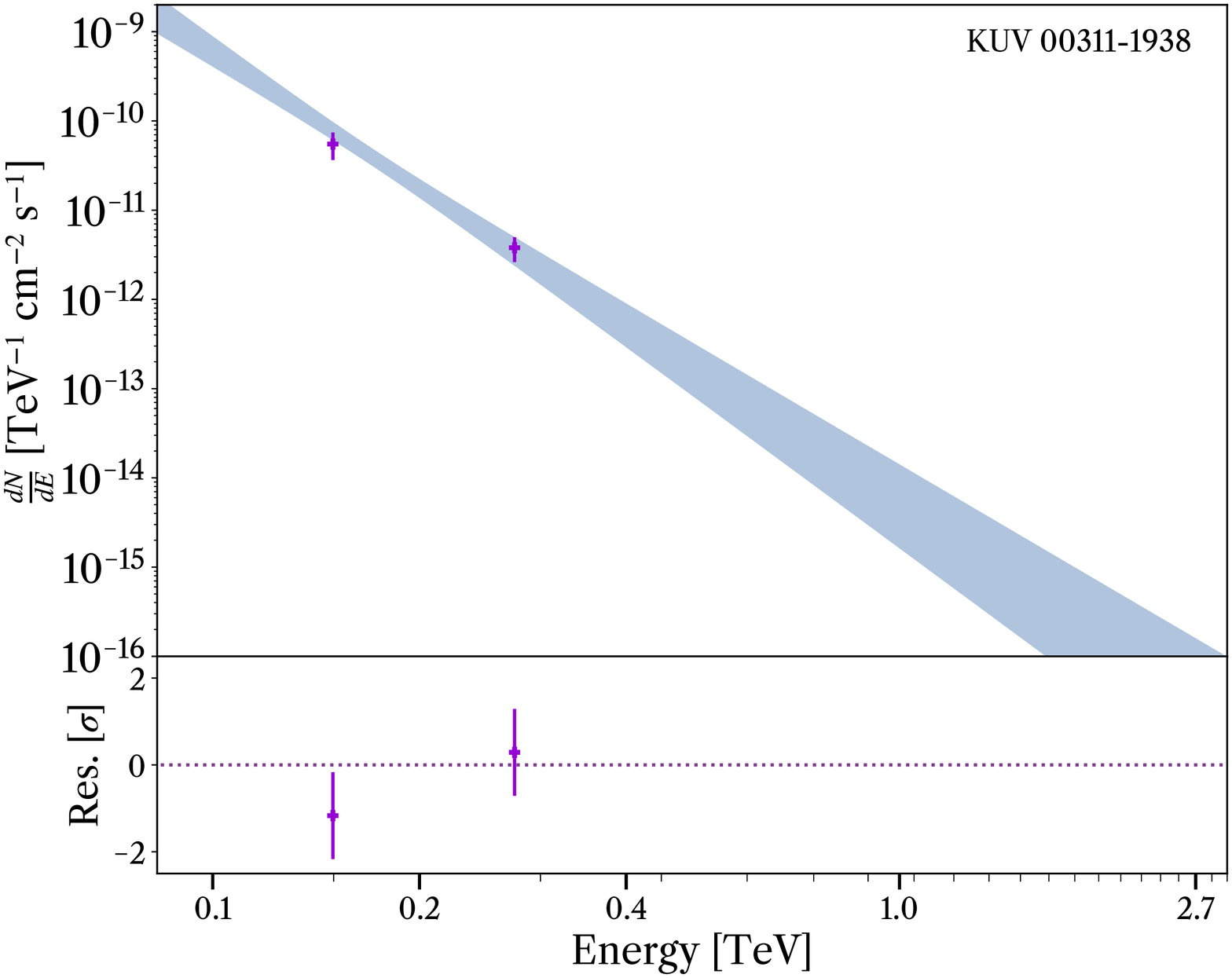}
  \hspace{-.94cm}
  \includegraphics[width=1.11\columnwidth]{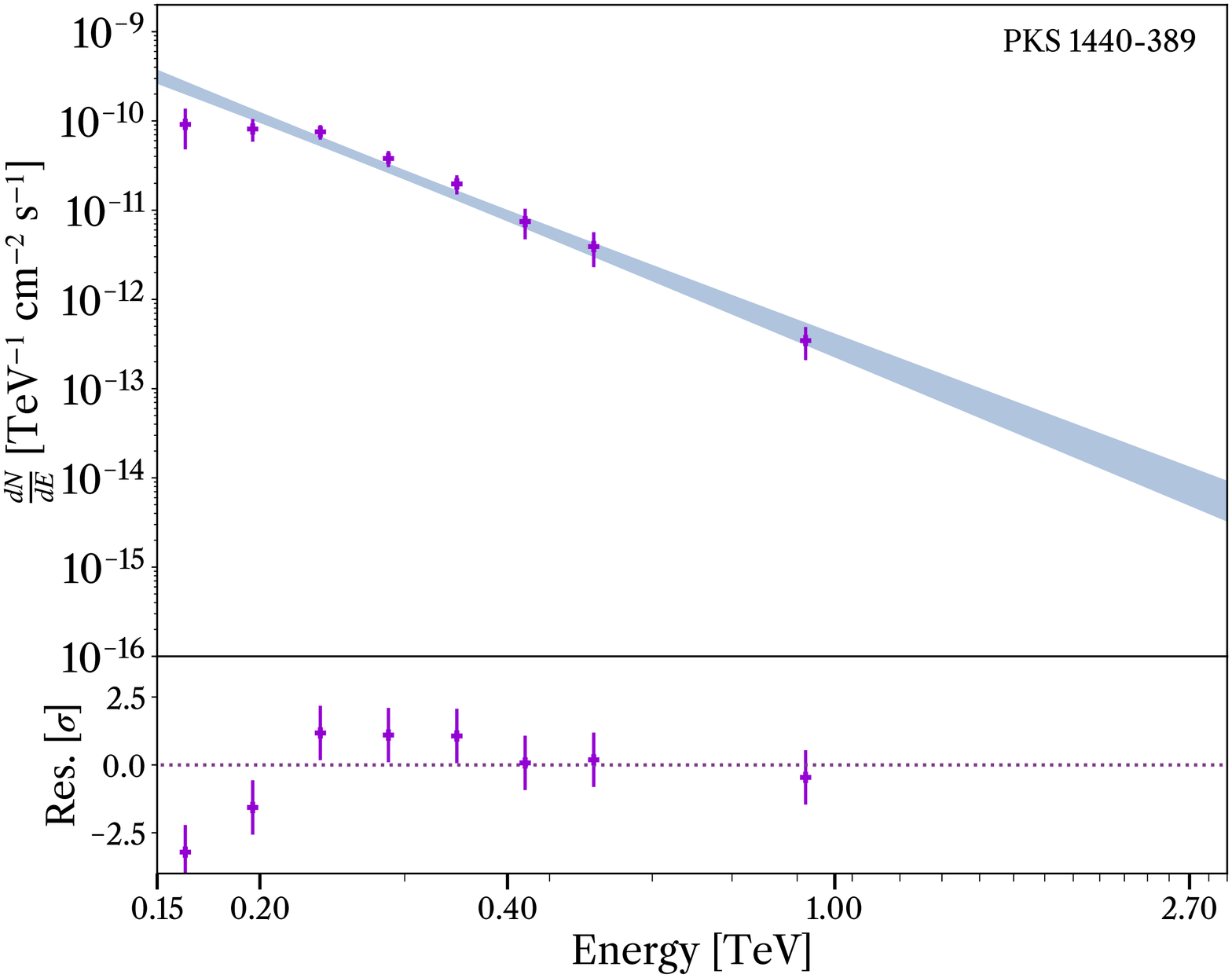}
  \caption{
	 \textsc{Left panel}:
 	 Differential energy spectrum of the VHE \gray\ emission of \kuv. 
	 \textsc{Right panel}:
	 Differential energy spectrum of the VHE \gray\ emission of \pks.   
    \textit{Upper plot}:
    Time-averaged VHE spectrum measured from the two sources.
    Overlaid spectral points were rebinned, requiring a minimum point significance of 2$\sigma$ per bin.
    The butterfly represents the $1 \sigma$ confidence level error band of the fitted spectrum using a power-law hypothesis.
	\textit{Lower plot}:
	Residuals of the reconstructed data points compared to the model.
  }
  \label{fig:kuvspectrum}\label{fig:pksspectrum}
\end{figure*}

The photon spectrum in the energy range above 147 \,GeV (shown in Fig. \ref{fig:pksspectrum}), and the fit results are found in \autoref{tab:hessfit}. The systematics were estimated in the same way as for the analysis of \kuv.

All the results have been cross-checked and confirmed with the analysis method in \citet{Becherini11}.

\section{Data analysis and results of multi-wavelength instruments}\label{multi}

Complementary to the H.E.S.S. observations, multi-wavelength data from observations  with \fermi\ (20~MeV--300~GeV), \swiftxrt\ (0.2--10~keV), \swiftuvot\ (170--650~nm) and \atom\  (optical R filter) are presented in this section. 
Only for \pks\ are all data contemporaneous. 
Unfortunately, for \kuv\ no contemporaneous \swift\ observations are available.

\subsection{\fermi}\label{fermigeneral}
The Large Area Telescope (LAT) on board the {\it Fermi} satellite is a pair-conversion \gray\ detector, sensitive in the energy range from  20~MeV to above 300~GeV \citep{Atwood09}. 

Data contemporaneous to the H.E.S.S. observations were analysed with the LAT ScienceTools, version {\tt v11r5p3} for both sources. 
Source-class events in a circular region of interest of $10^\circ$ radius centred at the positions of the sources are considered and the \texttt{P8R2\_SOURCE\_V6} instrumental response functions were used. 
To remove \graynoun s produced by cosmic-ray interactions in the Earth's limb, events with zenith angles greater than 90$^\circ$ were rejected. 
The isotropic background, containing both the extragalactic diffuse \g-ray and residual instrumental background, is estimated through the \texttt{iso\_p8v2\_SOURCE\_V6\_v06} model, while the Galactic diffuse emission is modelled with the spatial template \texttt{gll\_iem\_v06}. 
Spectral parameters were extracted by fitting a model containing the diffuse background and point sources from the 3FGL catalogue \citep[][]{3fgl}. 
The spectral parameters of sources within the inner 3$^\circ$ of the region of interest were left free during the fit, all others were fixed to their 3FGL values. 

The spectral analysis was performed between different values of $\rm E_{min}$ and $\rm E_{max}$ assuming three different spectral models, a power-law (PWL), a power-law with exponential cutoff (EC) or a LogParabola (LP), see Tab \ref{tab:latshapes} for full expressions.

\begin{table*}
  \centering
  \caption{Spectral models used in the \hess\ and \fermi\ analyses.}
  \label{tab:latshapes}
  \begin{tabular}{p{4cm} p{4.5cm} p{6cm}}
    \hline \hline
Spectral shape & Formula & Parameter explanation \\    
    \hline \hline
Power-law (PWL) & $\rm dN/dE = {N_0 \cdot ( E/E_0 )^{-\Gamma}}$ & $N_0$= normalisation at $E=E_0$, $\Gamma=$ photon index. $N_0$ and $\Gamma$ free in the minimisation \\
\hline
Power-law with exponential cutoff (EC) & $\rm dN/dE = N_0 \cdot (E/E_0)^{-\Gamma}\cdot exp{(-E/E_c)}$ & $E_c=$ cut-off energy, $N_0$ and $\Gamma$ as above. $N_0$, $\Gamma$, $E_c$ free in the minimisation \\
\hline
Log-parabola (LP) &  $\rm dN/dE = N_0 \cdot (E/E_b)^{-a-b\cdot\mathrm{log}(E/E_b)}$ & $E_b=$ scale parameter, $a=$ photon index at $E=E_b$, and $b=$ curvature parameter. $N_0$, $a$ and $b$ free in the minimisation \\
   \hline 
  \end{tabular}
\end{table*}

The three spectral models were used to assess the best spectral fit for a given analysis using the log likelihood ratio test. 
Results of the preferred spectral fits are summarised in \autoref{tab:lat}.
The event analysis presented here uses the binned likelihood method \citep{2009ApJ...697.1071A} with Pass8 (version 2) data and the user contributed python tools \texttt{Enrico} \citep{enrico}. 
The estimated systematic uncertainty on the flux in these analyses is 10\% at 0.1 GeV, 5\% at 0.5 GeV and 10\% at 10 GeV and above \citep{2011ApJ...743..171A}. 
The test positions of the sources were taken from the 3FGL and are consistent with the nominal positions given previously.

For each source, we present the analysis up to $\rm E_{max}= 300 \; GeV$ and the analysis up to $\rm E_{max}= 10 \; GeV$. 
The former analysis is used to understand the agreement with the \hess\ spectrum in the SED shown in Section 4.1.
The latter is used to assess the shape of the spectrum up at the energies where EBL absorption effects are negligible, so that the upper limit on the redshift can be evaluated, see Section 4.2.

\begin{table*}
  \centering
  \caption{Results of the spectral fitting of the \fermi\ data for the blazars \kuv\ and \pks. Only the most significant spectral shape of the various analyses presented, evaluated with the log-likelihood ratio test, is shown. We present the values of all the spectral parameters which were left free during the fit procedure.  $E_0$ is set to be the decorrelation energy of the fit. }
  \label{tab:lat}
  \begin{tabular}{lc cc lc}
    \hline \hline
Source & Model & Energy range & TS & Spectral parameters & Integrated flux \\
& &[GeV]&&&[$\rm ph\,cm^{-2} s^{-1}$]\\
\hline \hline
\kuv\ & EC  &$0.1-300$& \FermiPeriodECTS & $\Gamma = \FermiPeriodECIndex$,  & $F_{(E>100\,\mathrm{MeV})} = \FermiPeriodECIflux $\\ 
 &   & &  & $\rm N_{0} = \FermiPeriodECNorm $,  & \\ 
 &   & &  & $\rm E_{c} = \FermiPeriodECCut $  & \\ 
 &   & &  & $\rm E_{0} = 100 ~\rm MeV $  & \\ 
\kuv\ & PWL &$0.1-10$& 1465 & $\Gamma = 1.60 \pm 0.05$  & $F_{(E>100\,\mathrm{MeV})} = (1.1\pm 0.1)\times10^{-8}$\\
&   & &  & $\rm N_{0} = (4.46 \pm 0.23)\times10^{-13} \rm ph/MeV/cm^{2}/s$,  & \\ 
 &   & &  & $\rm E_{0} = 2398 ~\rm MeV $  & \\ 

\hline
\pks & PWL &$0.3-300$ & 88.4 & $\Gamma = 1.69\pm 0.15$ & $F_{(E>300\,\mathrm{MeV})} = (6.5\pm1.8)\times10^{-9}$\\ 
&   & &  & $\rm N_{0} = (3.57 \pm 0.79)\times10^{-13} \rm ph/MeV/cm^{2}/s$,  & \\ 
 &   & &  & $\rm E_{0} = 2753 ~\rm MeV $  & \\ 
\pks & PWL & $0.3-10$ & 57.5 & $\Gamma = 1.46\pm 0.30$ & $F_{(E>300\,\mathrm{MeV})} = (5.4\pm1.9)\times10^{-9}$\\ 
&   & &  & $\rm N_{0} = (4.04 \pm 0.98)\times10^{-13} \rm ph/MeV/cm^{2}/s$,  & \\ 
&   & &  & $\rm E_{0} = 2753 ~\rm MeV $  & \\ 
\hline
  \end{tabular}
\end{table*}

\subsubsection{\kuv}\label{fermianalysis} 
\fermi\ data contemporaneous to the \hess\ observations, i.e. in the period from \FermiPeriodStartObs\ to \FermiPeriodEndObs\ were analysed in the \KuvFermiEnergyRange\ energy range and in the \KuvFermiEnergyRangeLow\ energy range.

As the source has modest variability in the 4FGL catalogue (Section~\ref{KUVintro}), a 2-month binning  \fermi\ light curve was computed assuming a PWL shape, leaving the index free to vary (see Fig. \ref{fig:LightCurve}). 

\KuvFermiEnergyRange. No variability is detected during the \hess\ observation period, as an excess variance calculation yields a value compatible with zero.
The \gray\ emission from \kuv\ is therefore well described as constant in the \fermi\ energy range, implying that it is safe to combine the data from the full \hess\ observation period into a single spectrum. 
All spectral models result in an excess with a significance of about \FermiPeriodSigma.
A likelihood ratio test shows that the EC ($E_c = 54.4$ GeV) is preferred to the simple PWL shape at the \FermiPeriodRatioSig\ $\sigma$-level using five years of data, in line with the  \ensuremath{5.15 \sigma} for the whole 8-years \fermi\ 4FGL analysis. 
The reconstructed \fermi\ spectrum is shown alongside the \hess\ spectrum in Fig.\ \ref{fullSED}. The significance of the \fermi\ binned spectral points shown  is at least $3 \sigma$. If the significance of the bin is less than $3 \sigma$, a 95 \% upper limit on the flux in the bin is computed. 

\KuvFermiEnergyRangeLow. All spectral models result in an excess with a significance of about \KuvLowFermiPeriodSigma.
A likelihood ratio test shows that the PWL is the preferred shape in this energy range using the five years of data. 

\subsubsection{\pks}\label{LAT}
\fermi\ data analysis has been performed for the 3-month data set contemporaneous with the \hess\ observations (MJD~$55985-56074$) in the energy range \PksFermiEnergyRange\ and \PksFermiEnergyRangeLow. The low energy bound of 300~MeV was applied to avoid contamination from the bright, nearby quasar PKS~B1424$-$418 due to the larger point spread function at low energies \citep{2013ApJ...765...54A}.

Since \pks\ has modest variability in the 4FGL catalogue (Section~\ref{PKSintro}), a monthly binning  \fermi\ light curve was computed assuming a PWL shape, leaving the index free to vary (see Fig. \ref{fig:LightCurve}).   
The excess variance calculation over the \fermi\ lightcurve gives a value of $0.1\pm 0.3$. 
Therefore data from the whole 3-month dataset were combined into a single spectrum. 

\PksFermiEnergyRange. All spectral models in this energy range result in an excess with a significance of about \PksFermiPeriodSigma\ and the favoured shape is the PWL. 
The reconstructed \fermi\ spectrum in this range is shown alongside the \hess\ spectrum in Fig.\ \ref{fullSED}.

\PksFermiEnergyRangeLow. Spectral models in this energy range result in an excess with a significance of about $7.6\kern0.4pt\sigma$ and the preferred spectral shape is the PWL. 

\subsection{\swiftxrt\ and UVOT}

The X-ray telescope (XRT) on board the \swift\ satellite is designed to measure X-rays in the 0.2--10~keV energy range \citep{burrows2005}.\, Images in six filters (V and B in optical and U, UVW1, UVM2 and UVW2, in the ultra-violet, in order of increasing frequency) can be obtained simultaneously to XRT with the \swiftuvot\ telescope \citep{Roming05}.

Target of opportunity observations were obtained on 2012 April~29 (MJD~56046), following the \hess\ detection of \pks. 
Unfortunately that was not the case for \kuv\ and only a few observations exist, with none of these archival observations being within the time span of the overall \hess\ observing campaign. 

The X-ray observations were performed with the XRT detector in photon counting (PC) mode in the 0.3--10~keV energy range. 
The analysis was performed using the standard \texttt{HEASoft} (v6.16) and \texttt{Xspec} (v12.8.2) tools. 
Source counts were extracted using the {\tt xselect} task from a circular region with a radius of 20 pixels ($\sim 47$~arcsec).
Background counts were extracted from a 60-pixel circular region with no known X-ray sources.
The data were grouped, requiring a minimum of 20 counts per bin and then fitted with a power-law model including photo-electric absorption with a fixed value for the Galactic column density

Sky-corrected images for all available \swiftuvot\ filters were taken from the \swift\ archive, and aperture photometry was performed using the UVOT tasks included within the \texttt{HEASoft} package.  
Source counts were extracted using a $5\arcsec$ radius for all single exposures and all filters, while the background was estimated from different positions more than $25 \arcsec$ away from the source. 
Count rates were then converted to fluxes using the standard photometric zero-points \citep{Poole2008}. 
The reported fluxes are de-reddened for Galactic extinction following the procedure in \cite{Roming09}, with E(B-V) estimate from the IRSA\footnote{https://irsa.ipac.caltech.edu/applications/DUST/}.

\subsubsection{\kuv}\label{swiftx}

\begin{table*}
  \centering
  \caption{
  Available \swiftxrt\ observations, IDs, corresponding dates, and exposure times for \kuv\ (rows 1 to 6) and \pks\ (last row).
  Power-law parameters describing the differential photon flux obtained for the available XRT observations. $N_{\rm H \rm}$ is fixed at the Galactic value, $1.67\times10^{20}$~cm$^{-2}$ for \kuv\ and $1.08\times10^{21}$~cm$^{-2}$ for \pks.
  }
  \begin{tabular}{cccccc}
  \hline \hline
     ID & Start &  Exposure & $\Gamma$ & Normalisation at 1 keV &$\chi^2 /\rm d.o.f.$\\
        &       &   [s]      &          & [$10^{-3}$\,cm$^{-2}$\,s$^{-1}$\,keV$^{-1}$]& \\
    \hline \hline
 $00037546001$ &  $2008$-$11$-$09$ &  $1655$ & $2.13^{+0.15}_{-0.15}$ & $1.95^{+0.21}_{-0.18}$ &8.6/10 \\[1.2mm]
 $00037546002$ &  $2008$-$11$-$11$ &  $2922$ & $2.29^{+0.11}_{-0.11}$ & $1.76^{+0.12}_{-0.12}$ &22.1/26\\[1.2mm]
 $00037546003$ &  $2008$-$11$-$13$ &  $2790$ & $2.26^{+0.12}_{-0.12}$ & $1.60^{+0.17}_{-0.18}$ &23.4/23\\[1.2mm]
 $00037546004$ &  $2008$-$11$-$15$ &  $3891$ & $2.37^{+0.10}_{-0.10}$ & $1.61^{+0.15}_{-0.15}$ &41.3/32\\[1.2mm]
 $00038359001$ &  $2009$-$02$-$01$ &  $3349$ & $2.52^{+0.20}_{-0.21}$ & $0.86^{+0.08}_{-0.08}$ &16.2/15\\[1.2mm]
 $00038359002$ &  $2009$-$05$-$08$ &  $4722$ & $2.23^{+0.07}_{-0.07}$ & $3.54^{+0.19}_{-0.19}$ &93.2/82\\[0.3mm]
     \hline \hline
 $00041665002$ &  $2012$-$04$-$29$ &  $8551$ & $2.64^{+0.05}_{-0.05}$ & $3.80^{+0.12}_{-0.12}$ &124.4/108\\
  \end{tabular}  
\label{table:SwiftKuv}      
\end{table*}

We analyzed all six \swift\ snapshot observations of \kuv\ that were unfortunately performed prior to the \hess\ campaign, between 2008, November 9 and 2009, May 8, see \autoref{table:SwiftKuv}.  

The XRT spectra were fitted with a single power-law model with Galactic absorption $N_{\rm H}$ fixed at  $1.67\times10^{20}$~cm$^{-2}$ \citep{Willingale13}. 
Results are given for each individual observation in \autoref{table:SwiftKuv}, showing that the source is  variable in the X-ray band. 

\begin{table*}
  \centering
  \caption{\swiftuvot\ photometric fluxes in $10^{-12}$ erg\,cm$^{-2}$\,s$^{-1}$.  \kuv, rows 1 to 6; \pks, last row.}
  \begin{tabular}{ ccccccc }
    \hline \hline
	  ID         & V              & B              & U             & UVM1          & UVM2           & UVW2 \\
  \hline \hline
  $00037546001$ & 5.73$\pm$0.25  & 6.62$\pm$0.18  & 6.15$\pm$0.18 & 6.07$\pm$0.17 & 7.0 $\pm$0.19 & 6.76$\pm$0.14  \\
  $00037546002$ & 5.96$\pm$0.19  & 6.84$\pm$0.14  & 6.35$\pm$0.13 & 6.40$\pm$0.13 & 7.03$\pm$0.16 & 6.71$\pm$0.03  \\
  $00037546003$ & 6.34$\pm$0.20  & 6.68$\pm$0.14  & 6.17$\pm$0.14 & 6.12$\pm$0.13 & 7.52$\pm$0.17 & 6.70$\pm$0.11  \\
  $00037546004$ & 5.82$\pm$0.16  & 6.68$\pm$0.10  & 6.22$\pm$0.11 & 6.50$\pm$0.10 & 6.81$\pm$0.14 & 6.98$\pm$0.80  \\
  $00038359001$ & 6.12$\pm$0.13  & 6.57$\pm$0.76  & 6.12$\pm$0.84 & 6.22$\pm$0.86 &      -        & 6.56$\pm$0.75  \\
  $00038359002$ & 6.00$\pm$0.15  & 7.52$\pm$0.10  & 6.73$\pm$0.10 & 7.60$\pm$0.10 & 8.53$\pm$0.13 & 8.32$\pm$0.76  \\
  \hline \hline
  $00041665002$ & 17.34$\pm$0.56 & 18.07$\pm$0.42 & 18.11$\pm$0.42& 14.79$\pm$0.17& 19.77$\pm$0.27& 17.30$\pm$0.36 \\
  \end{tabular}  
\label{table:Swift_UVOT_Kuv}   
\end{table*}

During each XRT pointing, multiple exposures were taken with the \swiftuvot\ instrument using all its filters.
All available filters in each observation were searched for variability with the {\tt uvotmaghist} tool.
Since no variability was observed in any filter, the multiple images within each filter were stacked prior to performing aperture photometry.
The reported fluxes in \autoref{table:Swift_UVOT_Kuv} are de-reddened for Galactic extinction with  $E(B-V) = 0.019$~mag. 

The source exhibits variability between different observations, reaching a maximum flux around 2009 May 8 in both X-rays and ultraviolet. 
The UVOT photometric points and XRT spectral points are shown in \autoref{fig:SEDfit}.

\subsubsection{\pks}\label{Swift}

The fit result of the XRT spectrum of the 8\,ks exposure obtained on 2012 April~29 (MJD~56046) for a fixed Galactic column density of $N_{\mathrm{H}} = 1.08 \times 10^{21}$\,cm$^{-2}$ \citep{Willingale13}, can be found in \autoref{table:SwiftKuv}.

Results from the analysis of simultaneous \swiftuvot\ observations in all six filters are given in \autoref{table:Swift_UVOT_Kuv}. 
With UVW1 and UVM2 filter observations having four individual exposures, they were stacked prior to aperture photometry as the \texttt{uvotmaghist} task showed no variability between the individual exposures. 
The reported fluxes are de-reddened for Galactic absorption, $E(B-V) = 0.103$~mag.

\subsection{ATOM}
The Automatic Telescope for Optical Monitoring\footnote{See https://www.lsw.uni-heidelberg.de/projects/hess/ATOM/} (ATOM) is a $\rm 75\,cm$ optical support telescope for \hess, located at the \hess\ site. 
Operating since 2005, it provides optical monitoring of \g-ray sources. 
\kuv\  was regularly monitored from  April 2008 to December 2012 in the R band ($\rm 640\,nm$), while \pks\ was observed with high cadence during the whole 2012 \hess\ campaign on the source. 

The ATOM data were analysed using aperture photometry with custom calibrators in the field of view. We were able to use up to four calibrators in the case of \kuv\  and five for \pks. The resulting magnitudes were de-reddened in the same manner as done for the UVOT data. The uncertainty on the absolute calibration is propagated into the evaluation of the flux errors.
 
The resulting R-band light curves, corrected for Galactic extinction,  are presented in the bottom panel of \reffig{fig:LightCurve} showing clear variability in the optical regime contemporaneous with the \hess\ observations, for both \kuv\ and \pks.

\section{Discussion}\label{discussion}

In this section, we present a brief discussion of the implications of our observational results. 
Specifically, considering the unknown redshift of the sources, we will derive an upper limit on the redshift based on the extrapolation of the \fermi\ spectrum into the VHE (\hess) regime (Section~\ref{redshift}). 

\begin{figure*}
\includegraphics[width=15cm]{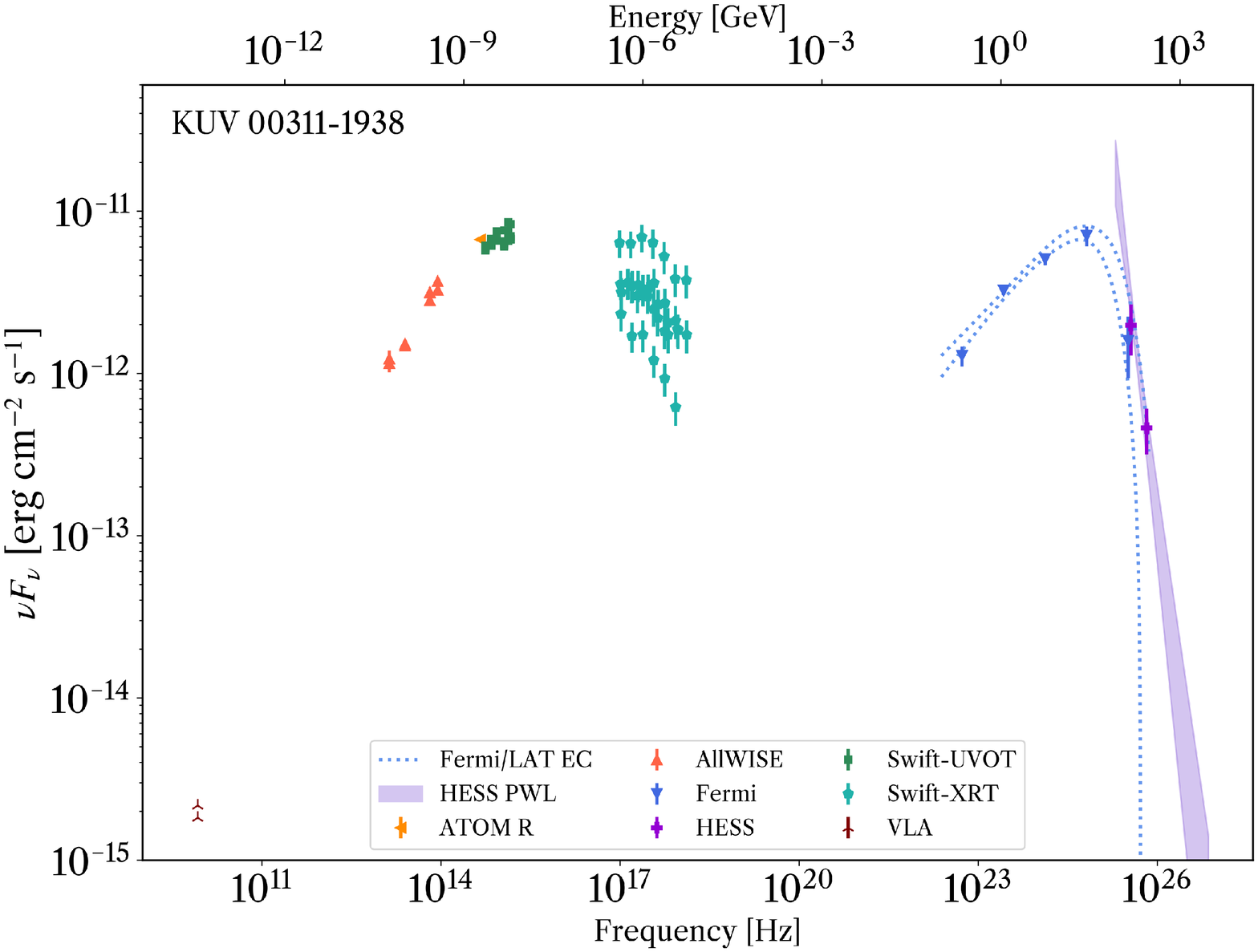}
\includegraphics[width=15cm]{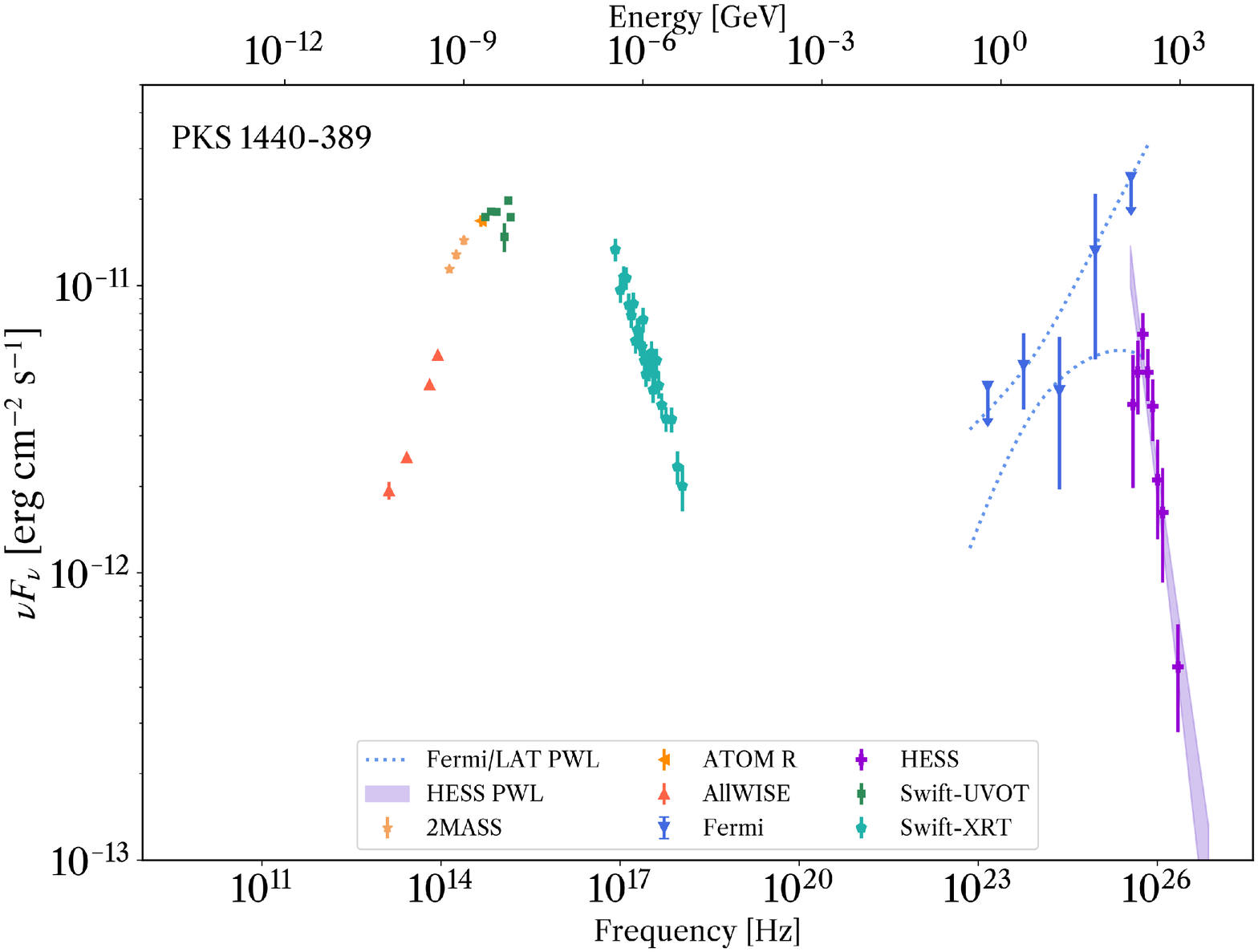}
  \caption{
  \textsc{Top Panel}:
    Averaged multi-wavelength spectral energy distribution of \kuv.
  \textsc{Bottom Panel}:
    Contemporaneous multi-wavelength spectral energy distribution of \pks.
    In both plots the \hess\ spectrum is represented by the filled butterfly and purple points at the highest energies, indicating 1$\sigma$ statistical errors.
    The measured \fermi\ spectrum is represented by the blue dotted lines and points (see Sec.\ \ref{fermianalysis} for details).
    Upper limits are calculated if the significance of the energy bin is less than 3 $\sigma$.
    The \fermi\ butterflies include only statistical errors.
    }
  \label{fig:SEDfit}\label{fullSED}
\end{figure*}

\subsection{Spectral Energy Distribution }\label{sed}
Here we present the broad band SED of the two sources, using both the observations presented above complemented by archival data at longer wavelengths and discuss them.

\subsubsection{\kuv}
\label{S-E-D}

To find archival data to complement the observations mentioned in the previous section, the SSDC SEDBuilder tool\footnote{Available at https://tools.ssdc.asi.it/SED/} was used to retrieve AllWISE \citep{2010AJ....140.1868W} data from the time period from the start of the first \swift\ observations to the end of the \hess\ observations, a time period spanning from November 2008 to November 2014.
The optical light curve for \kuv\ is summarised as the average flux.
In addition, radio fluxes of the central object as measured by the VLA \citep{2014ApJ...797...25P} during the extended \hess\ period are also included. 

The resulting broadband SED of \kuv\ is shown in Fig.\ \ref{fullSED}, displaying the standard synchrotron and inverse Compton peaks of similar luminosity. 

Grouping the \swiftxrt\ observation by date defines three flux states, and fitting the synchrotron peak at each of these three states with a third order polynomial gives a peak frequency and a peak flux. 
The low flux state results in a peak flux of $\sim$ \KuvLowNuFnuMax\ erg $\textrm{cm}^{-2} \textrm{s}^{-1}$ at  $\sim$ \KuvLowNuMax, 
the intermediate flux state results in $\sim$ \KuvMidNuFnuMax\ erg $\textrm{cm}^{-2} \textrm{s}^{-1}$ at $\sim$ \KuvMidNuMax\,
and $\sim$ \KuvHigNuFnuMax\ erg $\textrm{cm}^{-2} \textrm{s}^{-1}$ 
at $\sim$ \KuvHigNuMax\ for the high flux state.

Despite the variability in the X-rays, it is clear that the synchrotron peak is wider than the inverse Compton peak, a strong indication that the high energy bump has suffered some form of absorption at the higher energies. 
This is attributed to the effect of $\gamma$-$\gamma$ pair-production on the EBL in the line of sight.
This type of absorption becomes more important with increasing distance, and in the next section we use \gray\ observations and a model for the EBL to constrain the redshift range compatible with our observations.
It is also possible to get a similar break in the inverse Compton part of the spectrum if  Klein-Nishina effects are important \citep{2005ApJ...621..285K}.

\subsubsection{\pks}
Using the multi-wavelength data sets described in detail in the previous sections, a SED of \pks\ was constructed and is shown in \reffig{fig:SEDfit}. 
In addition to the data described in the previous sections, archival data from the 2MASS \citep{Skrutskie06,Mao11} and WISE catalogues \citep{2010AJ....140.1868W} were used to complete the low-energy part of the SED, bearing in mind that those data are not contemporaneous with the rest of the SED.  

Fitting the synchrotron peak with a third order polynomial gives a peak frequency of \PKSNuMax\ with a corresponding peak flux of \PKSNuFnuMax\ erg $\textrm{cm}^{-2} \textrm{s}^{-1}$.

\begin{figure*}
  \hspace{-.4cm}
  \includegraphics[width=1.1\columnwidth]{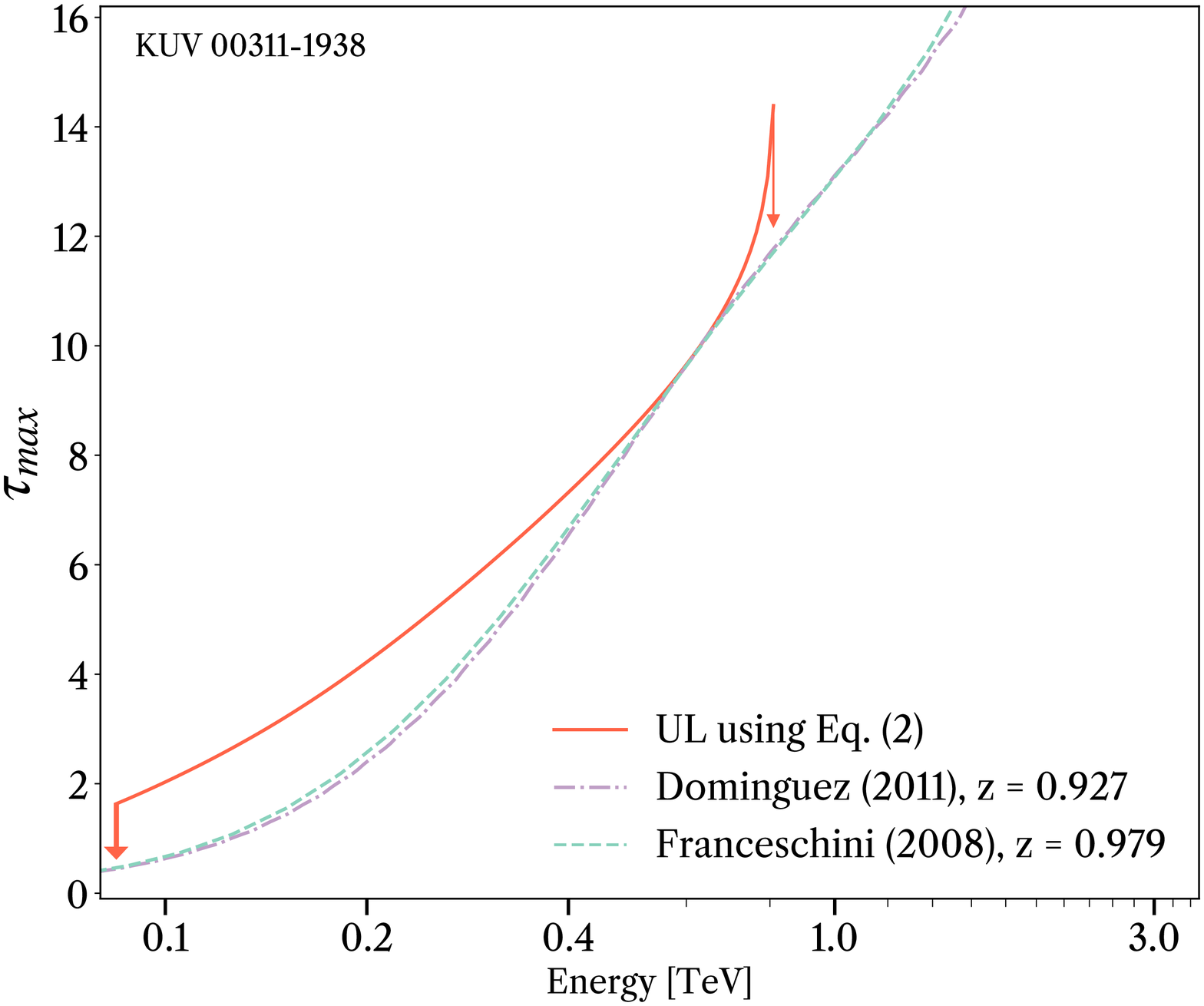}
  \hspace{-.9cm}
  \includegraphics[width=1.1\columnwidth]{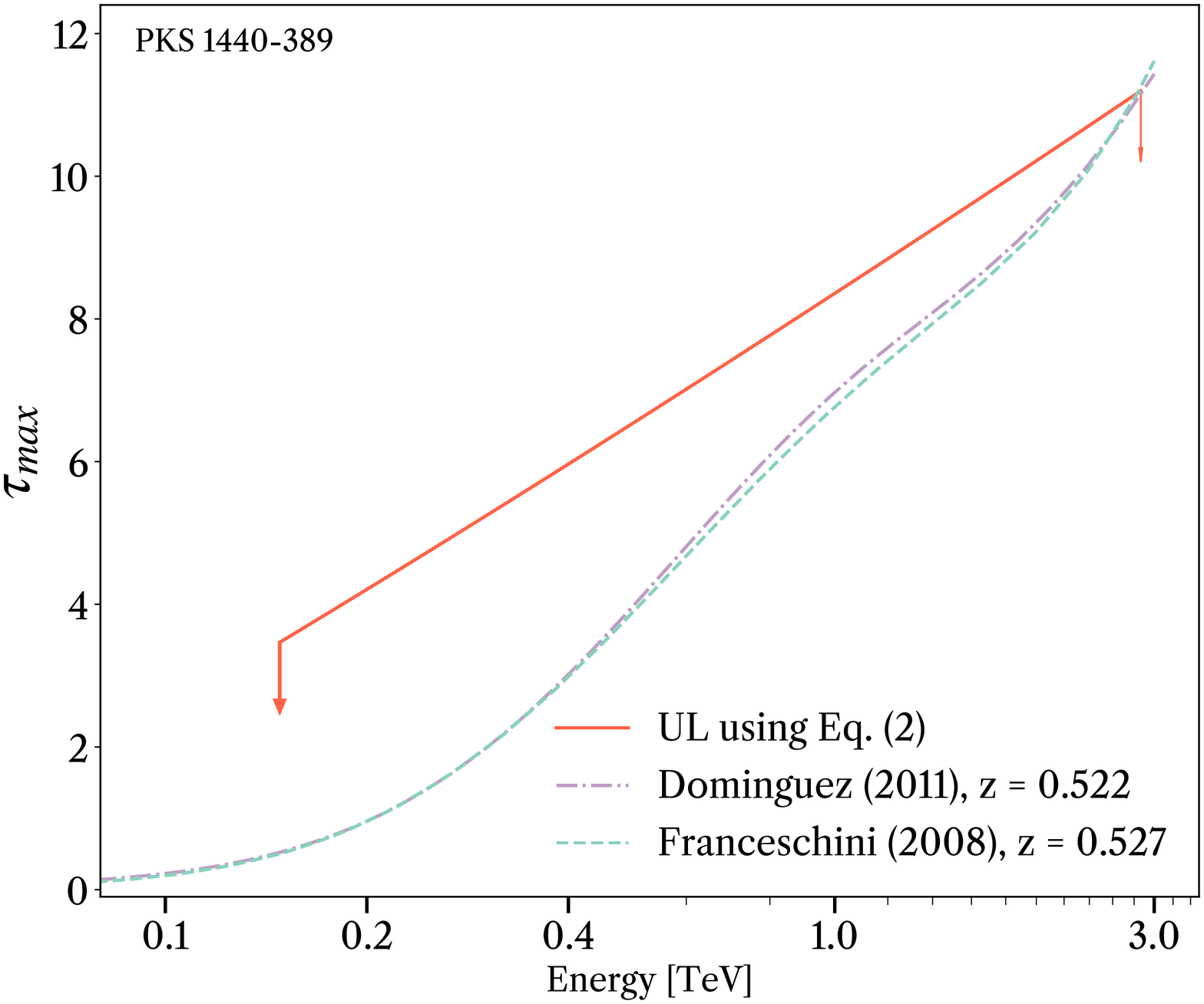}
  \caption{
    \textsc{ Left panel}: Optical depth as a function of energy for \kuv. 
    The red line represents the upper limit at the 95\% CL as derived from the combined \fermi\ and \hess\ data using \refeq{eq:tau2} which includes the \fermi\ errors. 
    Additionally shown by dashed and dot-dashed lines are the EBL model predictions from \citet{dom11} and \citet{fra08} that correspond to the upper limits on the redshift. \newline
    \textsc{ Right panel}: Optical depth as a function of energy for \pks. 
    The red line represents the upper limit at the 95\% CL as derived from the combined \fermi\ and \hess\ data using \refeq{eq:tau2}.
    Additionally shown by dashed and dot-dashed lines are the EBL model predictions for the two upper limits on the redshift. 
  }
  \label{fig:tau}
\end{figure*}

\subsection{Redshift constraints}\label{redshift}

The combined \fermi\ and \hess\ data set can be used to derive an upper limit on the redshift of the source, assuming that there is no upturn in the source-intrinsic \gray\ spectrum beyond the \fermi\ energy range. 
Under this assumption, the extrapolation of the HE \gray\ spectrum can be seen as an upper limit on the un-absorbed flux in the VHE regime. 
It therefore provides a conservative upper limit on the EBL absorption effect, which will be over-estimated if there is any downward curvature in the intrinsic \gray\ spectrum.

To obtain the intrinsic source spectrum, the energy range for the \fermi\ analysis was restricted to energies for which EBL absorption is negligible and then extrapolated to VHEs.
The ratio between this extrapolated flux, $F_\mathrm{{int}}(E)$, and the observed VHE flux, $F_\mathrm{{VHE}}(E)$, then provides an upper limit on the optical depth $\tau_{\mathrm{max}}(E)$, as $F_\mathrm{{VHE}}(E) / F_\mathrm{{int}}(E) =  \exp(-\tau_{\mathrm{max}}(E))$. 

Despite the small fraction of strictly overlapping HE and VHE observations, the \fermi\ spectra used for this calculation are a reasonable description of the intrinsic behaviour, because the lightcurves of both sources show only modest variability in the HESS observing periods, see \ref{fermianalysis} and \ref{LAT}.

Following Eq.~(1) from \citet{pks0447}, the upper limit on the optical depth at the (one sided) 95~\% confidence level (CL) can be written as
\begin{equation}
  \tau_\mathrm{{max}}(E) ~=~ \ln \left[ \frac{F_\mathrm{{int}}(E)} 
    {(1 - \eta_{\rm VHE}) \cdot ( F_\mathrm{{VHE}}(E) - 1.64 \cdot  \Delta
      F_\mathrm{{VHE}}(E))}  \right] \, 
\label{eq:tau}
\end{equation}
where $\Delta F_\mathrm{{VHE}}(E)$ denotes the statistical uncertainty of the VHE flux measurement and the term $(1-\eta_{\rm VHE})$ accounts for its systematic uncertainty. 
This is taken into account in a conservative way in the sense that $\eta_{\rm VHE}$ is the maximum factor by which the VHE flux could be overestimated.
For the presented \hess\ analyses, the factor $\eta_{\rm VHE}$ is \KuvFluxSysEta\ for \kuv\ and \PKSFluxSysEta\ for \pks, corresponding to the \KuvFluxSys\ and \PKSFluxSys\ systematic errors on the flux for \kuv\ and \pks\ respectively. Apart from the approximate treatment of the systematic error, \refeq{eq:tau} provides an exact expression for the 95\% CL upper limit on the optical depth if the intrinsic source spectrum is known precisely.

Including the statistical uncertainty of the \fermi\ spectrum in the evaluation of the confidence leads to the following modified version of \refeq{eq:tau}

\begin{equation}
  \tau_\mathrm{max}(E) ~=~ \ln \left[\frac{ a \cdot 
      (F_{\mathrm{int}}(E) + \frac{F_{\mathrm{VHE}}^{\star}}
      {(F_{\mathrm{int}}^{\star2}+F_{\mathrm{VHE}}^{\star2}-n^2)^{1/2}}  
      \;n \Delta F_{\mathrm{int}}(E))} 
    { b \cdot (F_{\mathrm{VHE}}(E) - \frac{F_{\mathrm{{int}}^{\star}}}
      {(F_{\mathrm{int}}^{\star2}+F_{\mathrm{VHE}}^{\star2}-n^2)^{1/2}} 
      \;n \Delta F_{\mathrm{VHE}}(E)) }  \right] \, 
\label{eq:tau2}
\end{equation}
with $a = (1+\eta_{\mathrm{HE}})$, $b=(1-\eta_{\mathrm{VHE}})$, $\Delta F_{\rm{int}}$ the uncertainty in the extrapolated flux, $F_i^{\star} \equiv F_i(E)/\Delta F_i(E)$  and $n=1.64$. 
The systematic uncertainty on the \fermi\ flux ($\eta_{\mathrm{HE}}$) has been added in a conservative way similar to the treatment of the \hess\ systematics  ($\eta_{\mathrm{VHE}}$) described above. 
We consider $\eta_{\mathrm{HE}}= 0.1$ for $E > 10~{\rm GeV}$, see also Section~\ref{fermigeneral}.
For full derivation of \refeq{eq:tau2} see \citet[][Appendix A]{pks0447}. 

Once an upper limit on the optical depth is calculated, using either \refeq{eq:tau} or  \refeq{eq:tau2}, the distance to a given blazar can be constrained by comparing it to the predictions of a specific EBL model. 
This is done by assuming a redshift, and then varying the assumed redshift until the difference between the predicted absorption and the observational upper limit is minimised.
The EBL models considered in this study are: the model of \citet{fra08} and the model of \citet{dom11}.

For the calculation of $\tau_{\rm max}$ below, we use \refeq{eq:tau2}.
We then compare the obtained $\tau_{\rm max}$ with the predictions coming from the two models.
We finally estimate a redshift by trying different source redshifts and then requiring that the EBL absorption by the models should be below the maximum permissible across the energy range.

\subsubsection{\kuv}
A power-law fitted to the \fermi\ data below 10~GeV (see \autoref{tab:lat}) and extrapolated to the \KuvEth\ to 3.0 TeV energy band was used as the intrinsic model, $F_{\rm int}$. This was then compared to the \hess\ observations via the measured $F_{\rm{VHE}}$ and $\Delta F_{\rm{VHE}}$.

We derive an upper limit of $z < \KuvUppzDom$ using \citet{dom11}. In comparison, using the Franceschini model we get a limit of $ z < \KuvUppzFra$.

The derived upper limit on the optical depth as a function of energy is shown on the left in \reffig{fig:tau}, calculated with \refeq{eq:tau2}. The \hess\ constraint together with the lower-limit from optical spectroscopy, limits the the redshift of \kuv\ to $\KuvLowz \leq z < \KuvUppz$ at the 95\% CL, improving on the previous upper limit of $z < 1.54 $.

\subsubsection{\pks}
To obtain the intrinsic source spectrum for \pks, the energy range for the \fermi\ analysis was restricted to energies below 10\,GeV. 
This was then extrapolated and used to estimate absorption in the energy range $\PksEth -3.0$~TeV. 

An upper limit of $z < \PKSUppz$ can be put on the redshift at the 95\% CL when using the EBL model of \citet{dom11}. Applying the Franceschini model yields a  compatible redshift constraint. This result significantly reduces the redshift constraint of $z<2.2$ obtained through optical spectroscopy by \cite{2013ApJ...764..135S}. 

Combining our constraint with the lower limit from the non-detection of the host galaxy by \cite{2013ApJ...764..135S}, the redshift of \pks\ is found to be in the range of $\PKSLowz \lessapprox z < \PKSUppz$ at the 95\% CL.

\section{Summary and Conclusions}\label{summary}
\label{END}

In this paper, we present the final H.E.S.S. results on the detection of the two blazars \kuv\ and \pks\ and their multi-wavelength analysis. With the shape of the inverse Compton bump we also constrain the redshift of the sources.

The blazar \kuv\ has been observed by \hess\ between 2009 and 2014, leading to the detection of its VHE emission with a significance of \KuvSig\ standard deviations.
The time-averaged VHE spectrum of this blazar is soft, with a photon index of \KuvIndex\ and a flux \KuvCrabatPKS\ of that of the Crab nebula above \PksEth.

Observations at other wavelengths have been analysed in order to have a multi-band view of the SED of this AGN detected in the VHE range. 
In particular, an analysis of the HE emission in \fermi\ data reveals a detection of this AGN at a significance level of about \FermiPeriodSigma\ standard deviations.

We reported also the detection of VHE \gray\ emission from the HSP \pks\ by \hess, accompanied by multi-wavelength observations with \fermi, \swiftxrt\ and UVOT, and \atom. 
The \hess\ spectrum has a spectral index of \PksIndexValue\ and a flux of  \PKSHessCrabFlux\ of that of the Crab nebula above \PksEth.
No evidence for short-term (day-scale or intra-day) variability has been found in the \gray\ regime (neither with \fermi\ nor with \hess), but moderate variability on weekly timescales is found in the optical light curve measured by \atom. 

By comparing the EBL absorption derived from models of intrinsic emission fitted to data, we constrain the redshift of \kuv\ to below $z = \KuvUppz $, which combines with known spectroscopical constraints to give $\KuvLowz \leq z < \KuvUppz$.
While VHE blazars have been detected up to $z=0.94$, the most distant ones have been seen only during flaring states. The current most distant and, within the limited duty cycle of IACTs, that are monitoring the VHE sky at the current sensitivity level only for the last two decades, persistent (since observations began and within observational constraints), VHE blazar is PKS~1424+240 \citep{2010ApJ...708L.100A} at $z\simeq0.6$. \kuv\ can thus potentially be the most distant persistent emitter of VHE photons. The determination of its redshift via optical spectroscopy is thus of paramount importance for VHE astrophysics: once its distance is constrained, it will be one of the best sources to study the propagation of VHE photons in the Universe.

For the blazar \pks\, based on an extrapolation of the \fermi\ spectrum towards the measured \hess\ VHE \gray\ spectrum, accounting for EBL absorption, we deduce an upper limit on the redshift of the source of $z < \PKSUppz$.
Combined with previous results, this constrains the redshift to the range $\PKSLowz \lessapprox  z < \PKSUppz$. 
\pks\ is found to be a standard source within the VHE HSPs, fitting reasonably well into the current population.

\section*{Acknowledgements}
The support of the Namibian authorities and of the University of Namibia in facilitating the construction and operation of H.E.S.S. is gratefully acknowledged, as is the support by the German Ministry for Education and Research (BMBF), the Max Planck Society, the German Research Foundation (DFG), the Helmholtz Association, the Alexander von Humboldt Foundation, the French Ministry of Higher Education, Research and Innovation, the Centre National de la Recherche Scientifique (CNRS/IN2P3 and CNRS/INSU), the Commissariat \`a l'\'energie atomique et aux \'energies alternatives (CEA), the U.K. Science and Technology Facilities Council (STFC), the Knut and Alice Wallenberg Foundation, the National Science Centre, Poland grant no. 2016/22/M/ST9/00382, the South African Department of Science and Technology and National Research Foundation, the University of Namibia, the National Commission on Research, Science and Technology of Namibia (NCRST), the Austrian Federal Ministry of Education, Science and Research and the Austrian Science Fund (FWF), the Australian Research Council (ARC), the Japan Society for the Promotion of Science and by the University of Amsterdam. We appreciate the excellent work of the technical support staff in Berlin, Zeuthen, Heidelberg, Palaiseau, Paris, Saclay, T\"ubingen and in Namibia in the construction and operation of the equipment.

This work benefited from services provided by the H.E.S.S. Virtual Organisation, supported by the national resource providers of the EGI Federation.

The H.E.S.S. and \fermi\ analysis computations were performed on resources provided by the Swedish National Infrastructure for Computing (SNIC) at Lunarc. 
Tomas Bylund and Yvonne Becherini wish to acknowledge the support of the Data Intensive Sciences and Applications (DISA) centre at Linnaeus University.

Matteo Cerruti has received financial support through the Postdoctoral Junior Leader Fellowship Programme from la Caixa Banking Foundation (LCF/BQ/PI18/11630012). 

This research made use of the NASA/IPAC Extragalactic Database (NED) and of the SIMBAD Astronomical Database, of data products from the Wide-field Infrared Survey Explorer, which is a joint project of the University of California, Los Angeles, and the Jet Propulsion Laboratory/California Institute of Technology, funded by the National Aeronautics and Space Administration.

We acknowledge the use of public data from the Swift data archive.

This research made use of Gammapy, a community-developed core Python package for gamma-ray astronomy \citep{2017ICRC...35..766D}. 

This research has made use of data and software provided by the Fermi Science Support Center, managed by the HEASARC at the Goddard Space Flight Center.

Part of this work is based on archival data, software or online services provided by the Space Science Data Center - ASI.

This research has made use of the VizieR catalogue access tool, CDS, Strasbourg, France (DOI: 10.26093/cds/vizier).
The original description of the VizieR service was published in A\&AS 143, 23.




\bibliographystyle{mnras}
\bibliography{References} 




\bsp	
\label{lastpage}
\end{document}

%% file: commands.tex
\newcommand{\pks}{PKS~1440$-$389}

\usepackage{xcolor}

\newcommand{\refeq}[1]{Eq.~(\ref{#1})}    
\newcommand{\reffig}[1]{Figure~\ref{#1}}  

\newcommand{\phcms}{\ensuremath{\rm ph\,cm^{-2}\,s^{-1} }}

\newcommand{\tevcms}{\ensuremath{\rm TeV^{-1}cm^{-2}s^{-1} }}

\newcommand{\cms}{{\rm cm}$^{-2}$\,s$^{-1}$}
\newcommand{\g}{\ensuremath{\gamma}}

\newcommand{\fermi}{\textsl{Fermi}/LAT}
\newcommand{\hess}{\textsc{H.E.S.S.}}
\newcommand{\atom}{\textsc{ATOM}}
\newcommand{\swift}{\textsl{Swift}}
\newcommand{\swiftxrt}{\textsl{Swift}/XRT}
\newcommand{\swiftuvot}{\textsl{Swift}/UVOT}
\newcommand{\kuv}{KUV~00311$-$1938}

\newcommand{\PksFermiEnergyRange}{[300~MeV--300~GeV]}
\newcommand{\PksFermiPeriodSigma}{\ensuremath{\approx 9.4 \sigma}}

\newcommand{\FermiFourFGLFlux}{$2.96 \pm 0.11\times 10^{-9}$ \phcms}

\newcommand{\FermiFourYearVar}{\ensuremath{ 0.16 \pm 0.06 }}
\newcommand{\FermiFourTwoMonthVar}{\ensuremath{ 0.33 \pm 0.05 }}

\newcommand{\FermiFourFGLPLIndex}{\ensuremath{1.67 \pm 0.02}}

\newcommand{\FermiFourFGLLPSig}{\ensuremath{4.68 \sigma }}

\newcommand{\FermiFourFGLECSig}{\ensuremath{5.15 \sigma }}

\newcommand{\FermiPKSFourFGLFlux}{$3.95 \pm 0.13\times 10^{-9}$ \phcms}

\newcommand{\FermiPKSFourYearVar}{\ensuremath{ 0.12 \pm 0.05 }}
\newcommand{\FermiPKSFourTwoMonthVar}{\ensuremath{ 0.16 \pm 0.05 }}

\newcommand{\FermiPKSFourFGLPLIndex}{\ensuremath{1.82 \pm 0.02}}

\newcommand{\PksFermiEnergyRangeLow}{[300~MeV--10~GeV]}

\newcommand{\FermiPKSFourFGLLPSig}{\ensuremath{4.14 \sigma }}

\newcommand{\FermiPKSFourFGLECSig}{\ensuremath{4.33 \sigma }}

\newcommand{\KuvFermiEnergyRange}{[100~MeV--300~GeV]}
\newcommand{\FermiPeriodStartObs}{2009-11-10 (MJD 55145)}
\newcommand{\FermiPeriodEndObs}{2014-08-25 (MJD 56894)}

\newcommand{\FermiPeriodECNorm}{\ensuremath{(6.99 \pm 1.05)\times 10^{-11} \rm ph/MeV/cm^{2}/s}}
\newcommand{\FermiPeriodECIndex}{\ensuremath{1.58 \pm 0.05}}
\newcommand{\FermiPeriodECCut}{\ensuremath{(54500 \pm 16200)}\; MeV}

\newcommand{\FermiPeriodECSRat}{\ensuremath{4.20}}
\newcommand{\FermiPeriodECTS}{\ensuremath{2009}}
\newcommand{\FermiPeriodECIflux}{\ensuremath{(1.15 \pm 0.1)\times 10^{-8}}}

\newcommand{\KuvFermiEnergyRangeLow}{[100~MeV--10~GeV]}

\newcommand{\FermiPeriodSigma}{\ensuremath{\approx 45 \sigma}}
\newcommand{\FermiPeriodRatioSig}{\FermiPeriodECSRat}
\newcommand{\KuvLowFermiPeriodSigma}{\ensuremath{\approx 7.6\kern0.4pt \sigma}}

\newcommand{\KuvTime}{102.6}
\newcommand{\KuvSig}{5.2}
\newcommand{\KuvOn}{4273}
\newcommand{\KuvOff}{53247}
\newcommand{\KuvAlpha}{0.073849}
\newcommand{\KuvExcess}{340.8}

\newcommand{\HessMeanZenit}{$14^\circ$}

\newcommand{\KuvEdec}{170 GeV}
\newcommand{\KuvEth}{83 GeV}
\newcommand{\KuvIndex}{\ensuremath{\Gamma = 5.1 \pm 0.6_{\rm stat} \pm 0.3_{\rm sys}}}
\newcommand{\KuvIndexValue}{\ensuremath{ 5.1 \pm 0.6_{\rm stat} \pm 0.3_{\rm sys}}}

\newcommand{\KuvFluxOverThreshValue}{\ensuremath{ 
			(3 \pm 1_{\rm stat} \pm 1_{\rm sys}) \times 10^{-11}} }

\newcommand{\KuvCrabatPKS}{\ensuremath{\sim 0.9\%}}
\newcommand{\KUVRAfitted}{\ensuremath{\rm \alpha_{J2000} = 00^{h} 33 ^{m} 36^{s} \pm 06 ^{s}_{\rm stat} }}
\newcommand{\KUVDECfitted}{\ensuremath{\rm \delta_{J2000} = -19^{\circ} 21' \pm 1'_{\rm stat}}}

\newcommand{\HESSRAsys}{\ensuremath{\rm \alpha_{J2000} =  \pm 1.3 ^{s}_{\rm sys}}}
\newcommand{\HESSDECsys}{\ensuremath{\rm \delta_{J2000} = \pm 20''_{\rm sys}}}

\newcommand{\RAVLA}{\ensuremath{\rm \alpha_{\rm J2000} = 00^{h} 33 ^{m} 34.380^{s} }\ }
\newcommand{\DECVLA}{\ensuremath{\rm \delta_{\rm J2000} = -19^{\circ} 21' 33.131''}\ }

\newcommand{\KuvFluxAtDecValue}{(4.1 \pm 0.8_{\rm stat} \pm  1.7_{\rm sys}) \times 10^{-11}}

\newcommand{\PksSig}{11.7}
\newcommand{\PksEdec}{274 GeV}
\newcommand{\PksEth}{147~{\rm GeV}}
\newcommand{\PksIndexValue}{\ensuremath{3.7 \pm 0.2_{\rm stat} \pm 0.3_{\rm sys}} }
\newcommand{\PksFluxOverThreshValue}{\ensuremath{ (1.87 \pm 0.21_{\rm stat} \pm 0.8_{\rm sys}) \times 10^{-11} }}
\newcommand{\PksFluxAtDecValue}{\ensuremath{(3.5 \pm 0.3_{\rm stat} \pm 1.2_{\rm sys}) \times 10^{-11} }}

\newcommand{\PKSHessCrabFlux}{\ensuremath{\sim 5.4\%}}

\newcommand{\PKSRAfitted}{\ensuremath{\rm \alpha_{\rm J2000} = 14 ^{h} 44 ^{m} 0.2 ^{s} \pm 2.6 ^{s}_{\rm stat} \pm 1.3 ^{s}_{\rm sys} }}
\newcommand{\PKSDECfitted}{\ensuremath{\rm \delta_{\rm J2000} = -39^{\circ} 08' 21'' \pm 31''_{\rm stat} \pm 20''_{\rm sys} }}

\newcommand{\PKSRAnom}{\ensuremath{\rm \alpha_{\rm J2000} = 14 ^{h} 43 ^{m} 57 ^{s}}}
\newcommand{\PKSDECnom}{\ensuremath{\rm \delta_{\rm J2000} = -39^{\circ} 08' 39''}}

\newcommand{\PKSFluxSysEta}{\ensuremath{0.36}}
\newcommand{\KuvFluxSysEta}{\ensuremath{0.42}}

\newcommand{\PKSFluxSys}{\ensuremath{36 \%}}
\newcommand{\KuvFluxSys}{\ensuremath{42 \%}}

\newcommand{\KuvLowNuMax}{\ensuremath{2 \times 10^{15}} Hz}
\newcommand{\KuvMidNuMax}{\ensuremath{3 \times 10^{15}} Hz}
\newcommand{\KuvHigNuMax}{\ensuremath{9 \times 10^{15}} Hz}
\newcommand{\KuvLowNuFnuMax}{\ensuremath{6.6 \times 10^{-12}} }
\newcommand{\KuvMidNuFnuMax}{\ensuremath{7 \times 10^{-12}} }
\newcommand{\KuvHigNuFnuMax}{\ensuremath{9 \times 10^{-12}} }

\newcommand{\PKSNuMax}{\ensuremath{4 \times 10^{15}} Hz}
\newcommand{\PKSNuFnuMax}{\ensuremath{2 \times 10^{-11}} }

\newcommand{\KuvLowz}{\ensuremath{0.51}}
\newcommand{\KuvUppz}{\KuvUppzFra}
\newcommand{\KuvUppzFra}{\ensuremath{0.98}}
\newcommand{\KuvUppzDom}{0.93}

\newcommand{\PKSLowz}{\ensuremath{0.14}}
\newcommand{\PKSUppz}{\ensuremath{0.53}}

%% file: authors.tex
\author[H.E.S.S. Collaboration]
{\Large\parbox{\textwidth}{
H.E.S.S. Collaboration,
H.~Abdalla ${}^{1}$,
R.~Adam ${}^{26}$,
F.~Aharonian ${}^{3,4,5}$,
F.~Ait~Benkhali ${}^{3}$,
E.O.~Ang\"uner ${}^{19}$,
M.~Arakawa ${}^{36}$,
C.~Arcaro ${}^{1}$,
C.~Armand ${}^{22}$,
T.~Armstrong ${}^{40}$
% ...
}}

%% file: affiliations.tex
\parbox{\textwidth}{
${}^{1}$Centre for Space Research, North-West University, Potchefstroom 2520, South Africa \\
${}^{2}$Universit\"at Hamburg, Institut f\"ur Experimentalphysik, Luruper Chaussee 149, D 22761 Hamburg, Germany \\
${}^{3}$Max-Planck-Institut f\"ur Kernphysik, P.O. Box 103980, D 69029 Heidelberg, Germany \\
${}^{4}$Dublin Institute for Advanced Studies, 31 Fitzwilliam Place, Dublin 2, Ireland \\
${}^{5}$High Energy Astrophysics Laboratory, RAU,  123 Hovsep Emin St  Yerevan 0051, Armenia \\
${}^{6}$Yerevan Physics Institute, 2 Alikhanian Brothers St., 375036 Yerevan, Armenia \\
${}^{7}$Institut f\"ur Physik, Humboldt-Universit\"at zu Berlin, Newtonstr. 15, D 12489 Berlin, Germany \\
${}^{8}$University of Namibia, Department of Physics, Private Bag 13301, Windhoek, Namibia, 12010 \\
${}^{9}$GRAPPA, Anton Pannekoek Institute for Astronomy, University of Amsterdam,  Science Park 904, 1098 XH Amsterdam, The Netherlands \\
${}^{10}$Department of Physics and Electrical Engineering, Linnaeus University,  351 95 V\"axj\"o, Sweden \\
${}^{11}$Institut f\"ur Theoretische Physik, Lehrstuhl IV: Weltraum und Astrophysik, Ruhr-Universit\"at Bochum, D 44780 Bochum, Germany \\
${}^{12}$Institut f\"ur Astro- und Teilchenphysik, Leopold-Franzens-Universit\"at Innsbruck, A-6020 Innsbruck, Austria \\
${}^{13}$School of Physical Sciences, University of Adelaide, Adelaide 5005, Australia \\
${}^{14}$LUTH, Observatoire de Paris, PSL Research University, CNRS, Universit\'e Paris Diderot, 5 Place Jules Janssen, 92190 Meudon, France \\
${}^{15}$Sorbonne Universit\'e, Universit\'e Paris Diderot, Sorbonne Paris Cit\'e, CNRS/IN2P3, Laboratoire de Physique Nucl\'eaire et de Hautes Energies, LPNHE, 4 Place Jussieu, F-75252 Paris, France \\
${}^{16}$Laboratoire Univers et Particules de Montpellier, Universit\'e Montpellier, CNRS/IN2P3,  CC 72, Place Eug\`ene Bataillon, F-34095 Montpellier Cedex 5, France \\
${}^{17}$IRFU, CEA, Universit\'e Paris-Saclay, F-91191 Gif-sur-Yvette, France \\
${}^{18}$Astronomical Observatory, The University of Warsaw, Al. Ujazdowskie 4, 00-478 Warsaw, Poland \\
${}^{19}$Aix Marseille Universit\'e, CNRS/IN2P3, CPPM, Marseille, France \\
${}^{20}$Instytut Fizyki J\c{a}drowej PAN, ul. Radzikowskiego 152, 31-342 Krak{\'o}w, Poland \\
${}^{21}$School of Physics, University of the Witwatersrand, 1 Jan Smuts Avenue, Braamfontein, Johannesburg, 2050 South Africa \\
${}^{22}$Laboratoire d'Annecy de Physique des Particules, Univ. Grenoble Alpes, Univ. Savoie Mont Blanc, CNRS, LAPP, 74000 Annecy, France \\
${}^{23}$Landessternwarte, Universit\"at Heidelberg, K\"onigstuhl, D 69117 Heidelberg, Germany \\
${}^{24}$Universit\'e Bordeaux, CNRS/IN2P3, Centre d'\'Etudes Nucl\'eaires de Bordeaux Gradignan, 33175 Gradignan, France \\
${}^{25}$Institut f\"ur Astronomie und Astrophysik, Universit\"at T\"ubingen, Sand 1, D 72076 T\"ubingen, Germany \\
${}^{26}$Laboratoire Leprince-Ringuet, École Polytechnique, CNRS, Institut Polytechnique de Paris, F-91128 Palaiseau, France \\
${}^{27}$Université de Paris, CNRS, Astroparticule et Cosmologie, F-75013 Paris, France \\
${}^{28}$Department of Physics and Astronomy, The University of Leicester, University Road, Leicester, LE1 7RH, United Kingdom \\
${}^{29}$Nicolaus Copernicus Astronomical Center, Polish Academy of Sciences, ul. Bartycka 18, 00-716 Warsaw, Poland \\
${}^{30}$Institut f\"ur Physik und Astronomie, Universit\"at Potsdam,  Karl-Liebknecht-Strasse 24/25, D 14476 Potsdam, Germany \\
${}^{31}$Friedrich-Alexander-Universit\"at Erlangen-N\"urnberg, Erlangen Centre for Astroparticle Physics, Erwin-Rommel-Str. 1, D 91058 Erlangen, Germany \\
${}^{32}$DESY, D-15738 Zeuthen, Germany \\
${}^{33}$Obserwatorium Astronomiczne, Uniwersytet Jagiello{\'n}ski, ul. Orla 171, 30-244 Krak{\'o}w, Poland \\
${}^{34}$Institute of Astronomy, Faculty of Physics, Astronomy and Informatics, Nicolaus Copernicus University,  Grudziadzka 5, 87-100 Torun, Poland \\
${}^{35}$Department of Physics, University of the Free State,  PO Box 339, Bloemfontein 9300, South Africa \\
${}^{36}$Department of Physics, Rikkyo University, 3-34-1 Nishi-Ikebukuro, Toshima-ku, Tokyo 171-8501, Japan \\
${}^{37}$Kavli Institute for the Physics and Mathematics of the Universe (WPI), The University of Tokyo Institutes for Advanced Study (UTIAS), The University of Tokyo, 5-1-5 Kashiwa-no-Ha, Kashiwa, Chiba, 277-8583, Japan \\
${}^{38}$Department of Physics, The University of Tokyo, 7-3-1 Hongo, Bunkyo-ku, Tokyo 113-0033, Japan \\
${}^{39}$RIKEN, 2-1 Hirosawa, Wako, Saitama 351-0198, Japan \\
${}^{40}$University of Oxford, Department of Physics, Denys Wilkinson Building, Keble Road, Oxford OX1 3RH, UK \\
${}^{41}$Institut de Ci\`{e}ncies del Cosmos (ICC UB), Universitat de Barcelona (IEEC-UB), Mart\'{i} Franqu\`es 1, E08028 Barcelona, Spain \\
}

%% file: Paper.bbl
\begin{thebibliography}{}
\makeatletter
\relax
\def\mn@urlcharsother{\let\do\@makeother \do\$\do\&\do\#\do\^\do\_\do\%\do\~}
\def\mn@doi{\begingroup\mn@urlcharsother \@ifnextchar [ {\mn@doi@}
  {\mn@doi@[]}}
\def\mn@doi@[#1]#2{\def\@tempa{#1}\ifx\@tempa\@empty \href
  {http://dx.doi.org/#2} {doi:#2}\else \href {http://dx.doi.org/#2} {#1}\fi
  \endgroup}
\def\mn@eprint#1#2{\mn@eprint@#1:#2::\@nil}
\def\mn@eprint@arXiv#1{\href {http://arxiv.org/abs/#1} {{\tt arXiv:#1}}}
\def\mn@eprint@dblp#1{\href {http://dblp.uni-trier.de/rec/bibtex/#1.xml}
  {dblp:#1}}
\def\mn@eprint@#1:#2:#3:#4\@nil{\def\@tempa {#1}\def\@tempb {#2}\def\@tempc
  {#3}\ifx \@tempc \@empty \let \@tempc \@tempb \let \@tempb \@tempa \fi \ifx
  \@tempb \@empty \def\@tempb {arXiv}\fi \@ifundefined
  {mn@eprint@\@tempb}{\@tempb:\@tempc}{\expandafter \expandafter \csname
  mn@eprint@\@tempb\endcsname \expandafter{\@tempc}}}

\bibitem[\protect\citeauthoryear{{Abdo} et~al.,}{{Abdo} et~al.}{2010}]{Abdo10}
{Abdo} A.~A.,  et~al., 2010, \mn@doi [\apj] {10.1088/0004-637X/716/1/30}, \href
  {http://adsabs.harvard.edu/abs/2010ApJ...716...30A} {716, 30}

\bibitem[\protect\citeauthoryear{{Abeysekara} et~al.,}{{Abeysekara}
  et~al.}{2019}]{VERITASEBL}
{Abeysekara} A.~U.,  et~al., 2019, \mn@doi [\apj] {10.3847/1538-4357/ab4817},
  \href {https://ui.adsabs.harvard.edu/abs/2019ApJ...885..150A} {885, 150}

\bibitem[\protect\citeauthoryear{{Acciari} et~al.,}{{Acciari}
  et~al.}{2010}]{2010ApJ...708L.100A}
{Acciari} V.~A.,  et~al., 2010, \mn@doi [\apjl] {10.1088/2041-8205/708/2/L100},
  \href {https://ui.adsabs.harvard.edu/abs/2010ApJ...708L.100A} {708, L100}

\bibitem[\protect\citeauthoryear{{Acciari} et~al.,}{{Acciari}
  et~al.}{2019}]{MAGICEBL}
{Acciari} V.~A.,  et~al., 2019, \mn@doi [\mnras] {10.1093/mnras/stz943}, \href
  {https://ui.adsabs.harvard.edu/abs/2019MNRAS.486.4233A} {486, 4233}

\bibitem[\protect\citeauthoryear{{Acero} et~al.,}{{Acero} et~al.}{2015a}]{3fgl}
{Acero} F.,  et~al., 2015a, \mn@doi [\apjs] {10.1088/0067-0049/218/2/23}, \href
  {http://adsabs.harvard.edu/abs/2015ApJS..218...23A} {218, 23}

\bibitem[\protect\citeauthoryear{{Acero} et~al.,}{{Acero}
  et~al.}{2015b}]{2015ApJS..218...23A}
{Acero} F.,  et~al., 2015b, \mn@doi [The Astrophysical Journal Supplement
  Series] {10.1088/0067-0049/218/2/23}, \href
  {https://ui.adsabs.harvard.edu/#abs/2015ApJS..218...23A} {218, 23}

\bibitem[\protect\citeauthoryear{{Ackermann} et~al.,}{{Ackermann}
  et~al.}{2011}]{2011ApJ...743..171A}
{Ackermann} M.,  et~al., 2011, \mn@doi [\apj] {10.1088/0004-637X/743/2/171},
  \href {https://ui.adsabs.harvard.edu/abs/2011ApJ...743..171A} {743, 171}

\bibitem[\protect\citeauthoryear{{Ackermann} et~al.,}{{Ackermann}
  et~al.}{2013}]{2013ApJ...765...54A}
{Ackermann} M.,  et~al., 2013, \mn@doi [\apj] {10.1088/0004-637X/765/1/54},
  \href {https://ui.adsabs.harvard.edu/abs/2013ApJ...765...54A} {765, 54}

\bibitem[\protect\citeauthoryear{{Aharonian} et~al.,}{{Aharonian}
  et~al.}{2006a}]{Aharonian06}
{Aharonian} F.,  et~al., 2006a, \mn@doi [\nat] {10.1038/nature04680}, \href
  {http://adsabs.harvard.edu/abs/2006Natur.440.1018A} {440, 1018}

\bibitem[\protect\citeauthoryear{{Aharonian} et~al.,}{{Aharonian}
  et~al.}{2006b}]{Aharonian06Crab}
{Aharonian} F.,  et~al., 2006b, \mn@doi [\aap] {10.1051/0004-6361:20065351},
  \href {http://adsabs.harvard.edu/abs/2006A%26A...457..899A} {457, 899}

\bibitem[\protect\citeauthoryear{{Aharonian} et~al.,}{{Aharonian}
  et~al.}{2007}]{Aharonian07}
{Aharonian} F.,  et~al., 2007, \mn@doi [\apjl] {10.1086/520635}, \href
  {http://adsabs.harvard.edu/abs/2007ApJ...664L..71A} {664, L71}

\bibitem[\protect\citeauthoryear{{Ahnen} et~al.,}{{Ahnen}
  et~al.}{2016}]{Ahnen16}
{Ahnen} M.~L.,  et~al., 2016, \mn@doi [\aap] {10.1051/0004-6361/201629461},
  \href {http://adsabs.harvard.edu/abs/2016A%26A...595A..98A} {595, A98}

\bibitem[\protect\citeauthoryear{{Ajello} et~al.,}{{Ajello}
  et~al.}{2017}]{2017ApJS..232...18A}
{Ajello} M.,  et~al., 2017, \mn@doi [\apjs] {10.3847/1538-4365/aa8221}, \href
  {http://adsabs.harvard.edu/abs/2017ApJS..232...18A} {232, 18}

\bibitem[\protect\citeauthoryear{{Albert} et~al.,}{{Albert}
  et~al.}{2007}]{Albert07}
{Albert} J.,  et~al., 2007, \mn@doi [\apj] {10.1086/521382}, \href
  {http://adsabs.harvard.edu/abs/2007ApJ...669..862A} {669, 862}

\bibitem[\protect\citeauthoryear{{Arlen} et~al.,}{{Arlen}
  et~al.}{2013}]{Arlen13}
{Arlen} T.,  et~al., 2013, \mn@doi [\apj] {10.1088/0004-637X/762/2/92}, \href
  {http://adsabs.harvard.edu/abs/2013ApJ...762...92A} {762, 92}

\bibitem[\protect\citeauthoryear{{Atwood} et~al.,}{{Atwood}
  et~al.}{2009a}]{2009ApJ...697.1071A}
{Atwood} W.~B.,  et~al., 2009a, \mn@doi [\apj] {10.1088/0004-637X/697/2/1071},
  \href {https://ui.adsabs.harvard.edu/#abs/2009ApJ...697.1071A} {697, 1071}

\bibitem[\protect\citeauthoryear{{Atwood} et~al.,}{{Atwood}
  et~al.}{2009b}]{Atwood09}
{Atwood} W.~B.,  et~al., 2009b, \mn@doi [\apj] {10.1088/0004-637X/697/2/1071},
  \href {http://adsabs.harvard.edu/abs/2009ApJ...697.1071A} {697, 1071}

\bibitem[\protect\citeauthoryear{{Bauer}, {Condon}, {Thuan}  \&
  {Broderick}}{{Bauer} et~al.}{2000}]{2000ApJS..129..547B}
{Bauer} F.~E.,  {Condon} J.~J.,  {Thuan} T.~X.,   {Broderick} J.~J.,  2000,
  \mn@doi [The Astrophysical Journal Supplement Series] {10.1086/313425}, \href
  {https://ui.adsabs.harvard.edu/#abs/2000ApJS..129..547B} {129, 547}

\bibitem[\protect\citeauthoryear{{Becherini}, {Djannati-Ata{\"i}}, {Marandon},
  {Punch}  \& {Pita}}{{Becherini} et~al.}{2011}]{Becherini11}
{Becherini} Y.,  {Djannati-Ata{\"i}} A.,  {Marandon} V.,  {Punch} M.,   {Pita}
  S.,  2011, \mn@doi [Astroparticle Physics]
  {10.1016/j.astropartphys.2011.03.005}, \href
  {http://adsabs.harvard.edu/abs/2011APh....34..858B} {34, 858}

\bibitem[\protect\citeauthoryear{{Becherini}, {Boisson}, {Cerruti}  \&
  {H.~E.~S.~S. Collaboration}}{{Becherini} et~al.}{2012}]{2012AIPC.1505..490B}
{Becherini} Y.,  {Boisson} C.,  {Cerruti} M.,   {H.~E.~S.~S. Collaboration}
  2012, in {Aharonian} F.~A.,  {Hofmann} W.,   {Rieger} F.~M.,  eds,  American
  Institute of Physics Conference Series Vol. 1505, American Institute of
  Physics Conference Series. pp 490--493, \mn@doi{10.1063/1.4772304}

\bibitem[\protect\citeauthoryear{{Bolmont} et~al.,}{{Bolmont}
  et~al.}{2014}]{2014NIMPA.761...46B}
{Bolmont} J.,  et~al., 2014, \mn@doi [Nuclear Instruments and Methods in
  Physics Research A] {10.1016/j.nima.2014.05.093}, \href
  {https://ui.adsabs.harvard.edu/\#abs/2014NIMPA.761...46B} {761, 46}

\bibitem[\protect\citeauthoryear{{B{\"o}ttcher}, {Reimer}, {Sweeney}  \&
  {Prakash}}{{B{\"o}ttcher} et~al.}{2013}]{Boettcher13}
{B{\"o}ttcher} M.,  {Reimer} A.,  {Sweeney} K.,   {Prakash} A.,  2013, \mn@doi
  [\apj] {10.1088/0004-637X/768/1/54}, \href
  {https://ui.adsabs.harvard.edu/abs/2013ApJ...768...54B} {768, 54}

\bibitem[\protect\citeauthoryear{{Burrows} et~al.,}{{Burrows}
  et~al.}{2005}]{burrows2005}
{Burrows} D.~N.,  et~al., 2005, \mn@doi [\ssr] {10.1007/s11214-005-5097-2},
  \href {http://adsabs.harvard.edu/abs/2005SSRv..120..165B} {120, 165}

\bibitem[\protect\citeauthoryear{{Condon}, {Cotton}, {Greisen}, {Yin},
  {Perley}, {Taylor}  \& {Broderick}}{{Condon}
  et~al.}{1998}]{1998AJ....115.1693C}
{Condon} J.~J.,  {Cotton} W.~D.,  {Greisen} E.~W.,  {Yin} Q.~F.,  {Perley}
  R.~A.,  {Taylor} G.~B.,   {Broderick} J.~J.,  1998, \mn@doi [\aj]
  {10.1086/300337}, \href
  {https://ui.adsabs.harvard.edu/#abs/1998AJ....115.1693C} {115, 1693}

\bibitem[\protect\citeauthoryear{{De Naurois} \& {Rolland}}{{De Naurois} \&
  {Rolland}}{2009}]{DNR09}
{De Naurois} M.,  {Rolland} L.,  2009, \mn@doi [Astroparticle Physics]
  {10.1016/j.astropartphys.2009.09.001}, \href
  {http://adsabs.harvard.edu/abs/2009APh....32..231D} {32, 231}

\bibitem[\protect\citeauthoryear{{Deil} et~al.,}{{Deil}
  et~al.}{2017}]{2017ICRC...35..766D}
{Deil} C.,  et~al., 2017, in 35th International Cosmic Ray Conference
  (ICRC2017). p.~766 (\mn@eprint {arXiv} {1709.01751})

\bibitem[\protect\citeauthoryear{{Dom{\'{\i}}nguez} et~al.,}{{Dom{\'{\i}}nguez}
  et~al.}{2011}]{dom11}
{Dom{\'{\i}}nguez} A.,  et~al., 2011, \mn@doi [\mnras]
  {10.1111/j.1365-2966.2010.17631.x}, \href
  {http://adsabs.harvard.edu/abs/2011MNRAS.410.2556D} {410, 2556}

\bibitem[\protect\citeauthoryear{{Fermi-LAT Collaboration} et~al.,}{{Fermi-LAT
  Collaboration} et~al.}{2018}]{FermiEBL}
{Fermi-LAT Collaboration} et~al., 2018, \mn@doi [Science]
  {10.1126/science.aat8123}, \href
  {https://ui.adsabs.harvard.edu/abs/2018Sci...362.1031F} {362, 1031}

\bibitem[\protect\citeauthoryear{{Fernandez Alonso}, {Pichel}  \&
  {Rovero}}{{Fernandez Alonso} et~al.}{2019}]{ICRC2019}
{Fernandez Alonso} M.,  {Pichel} A.,   {Rovero} A.,  2019, in Proceedings of
  Sciece.

\bibitem[\protect\citeauthoryear{{Finke}, {Razzaque}  \& {Dermer}}{{Finke}
  et~al.}{2010}]{Finke10}
{Finke} J.~D.,  {Razzaque} S.,   {Dermer} C.~D.,  2010, \mn@doi [\apj]
  {10.1088/0004-637X/712/1/238}, \href
  {http://adsabs.harvard.edu/abs/2010ApJ...712..238F} {712, 238}

\bibitem[\protect\citeauthoryear{{Fomin}, {Stepanian}, {Lamb}, {Lewis}, {Punch}
   \& {Weekes}}{{Fomin} et~al.}{1994}]{Fomin94}
{Fomin} V.~P.,  {Stepanian} A.~A.,  {Lamb} R.~C.,  {Lewis} D.~A.,  {Punch} M.,
   {Weekes} T.~C.,  1994, \mn@doi [Astroparticle Physics]
  {10.1016/0927-6505(94)90036-1}, \href
  {http://adsabs.harvard.edu/abs/1994APh.....2..137F} {2, 137}

\bibitem[\protect\citeauthoryear{{Franceschini}, {Rodighiero}  \&
  {Vaccari}}{{Franceschini} et~al.}{2008}]{fra08}
{Franceschini} A.,  {Rodighiero} G.,   {Vaccari} M.,  2008, \mn@doi [\aap]
  {10.1051/0004-6361:200809691}, \href
  {http://adsabs.harvard.edu/abs/2008A%26A...487..837F} {487, 837}

\bibitem[\protect\citeauthoryear{{Gaidos} et~al.,}{{Gaidos}
  et~al.}{1996}]{1996Natur.383..319G}
{Gaidos} J.~A.,  et~al., 1996, \mn@doi [\nat] {10.1038/383319a0}, \href
  {https://ui.adsabs.harvard.edu/abs/1996Natur.383..319G} {383, 319}

\bibitem[\protect\citeauthoryear{Gould \& Schr\'eder}{Gould \&
  Schr\'eder}{1967}]{GS67}
Gould R.~J.,  Schr\'eder G.~P.,  1967, \mn@doi [Phys. Rev.]
  {10.1103/PhysRev.155.1408}, 155, 1408

\bibitem[\protect\citeauthoryear{{H.E.S.S.~Collaboration}
  et~al.,}{{H.E.S.S.~Collaboration} et~al.}{2013}]{pks0447}
{H.E.S.S.~Collaboration} et~al., 2013, \mn@doi [\aap]
  {10.1051/0004-6361/201321108}, \href
  {http://adsabs.harvard.edu/abs/2013A%26A...552A.118H} {552, A118}

\bibitem[\protect\citeauthoryear{{H.E.S.S. Collaboration} et~al.,}{{H.E.S.S.
  Collaboration} et~al.}{2017}]{HESSEBL}
{H.E.S.S. Collaboration} et~al., 2017, \mn@doi [\aap]
  {10.1051/0004-6361/201731200}, \href
  {https://ui.adsabs.harvard.edu/abs/2017A&A...606A..59H} {606, A59}

\bibitem[\protect\citeauthoryear{{Jackson}, {Wall}, {Shaver}, {Kellermann},
  {Hook}  \& {Hawkins}}{{Jackson} et~al.}{2002}]{Jackson02}
{Jackson} C.~A.,  {Wall} J.~V.,  {Shaver} P.~A.,  {Kellermann} K.~I.,  {Hook}
  I.~M.,   {Hawkins} M.~R.~S.,  2002, \mn@doi [\aap]
  {10.1051/0004-6361:20020119}, \href
  {http://adsabs.harvard.edu/abs/2002A%26A...386...97J} {386, 97}

\bibitem[\protect\citeauthoryear{{Jones} et~al.,}{{Jones}
  et~al.}{2004}]{2004MNRAS.355..747J}
{Jones} D.~H.,  et~al., 2004, \mn@doi [\mnras]
  {10.1111/j.1365-2966.2004.08353.x}, \href
  {http://adsabs.harvard.edu/abs/2004MNRAS.355..747J} {355, 747}

\bibitem[\protect\citeauthoryear{{Jones} et~al.,}{{Jones}
  et~al.}{2009}]{2009MNRAS.399..683J}
{Jones} D.~H.,  et~al., 2009, \mn@doi [\mnras]
  {10.1111/j.1365-2966.2009.15338.x}, \href
  {http://adsabs.harvard.edu/abs/2009MNRAS.399..683J} {399, 683}

\bibitem[\protect\citeauthoryear{{Krau{\ss}} et~al.,}{{Krau{\ss}}
  et~al.}{2016}]{Krauss16}
{Krau{\ss}} F.,  et~al., 2016, \mn@doi [\aap] {10.1051/0004-6361/201628595},
  591, A130

\bibitem[\protect\citeauthoryear{{Krawczynski} et~al.,}{{Krawczynski}
  et~al.}{2004}]{Krawczynski04}
{Krawczynski} H.,  et~al., 2004, \mn@doi [\apj] {10.1086/380393}, \href
  {http://adsabs.harvard.edu/abs/2004ApJ...601..151K} {601, 151}

\bibitem[\protect\citeauthoryear{{Kusunose} \& {Takahara}}{{Kusunose} \&
  {Takahara}}{2005}]{2005ApJ...621..285K}
{Kusunose} M.,  {Takahara} F.,  2005, \mn@doi [\apj] {10.1086/427405}, \href
  {https://ui.adsabs.harvard.edu/abs/2005ApJ...621..285K} {621, 285}

\bibitem[\protect\citeauthoryear{{Landoni} et~al.,}{{Landoni}
  et~al.}{2015}]{2015AJ....149..163L}
{Landoni} M.,  et~al., 2015, \mn@doi [\aj] {10.1088/0004-6256/149/5/163}, \href
  {http://adsabs.harvard.edu/abs/2015AJ....149..163L} {149, 163}

\bibitem[\protect\citeauthoryear{{Li} \& {Ma}}{{Li} \& {Ma}}{1983}]{LiMa}
{Li} T.~P.,  {Ma} Y.~Q.,  1983, \mn@doi [\apj] {10.1086/161295}, \href
  {https://ui.adsabs.harvard.edu/#abs/1983ApJ...272..317L} {272, 317}

\bibitem[\protect\citeauthoryear{{Mao}}{{Mao}}{2011}]{Mao11}
{Mao} L.~S.,  2011, \mn@doi [\na] {10.1016/j.newast.2011.05.002}, \href
  {http://adsabs.harvard.edu/abs/2011NewA...16..503M} {16, 503}

\bibitem[\protect\citeauthoryear{{Nikishov}}{{Nikishov}}{1962}]{Nikishov62}
{Nikishov} A.~I.,  1962, Sov. Phys. JETP, 14, 393

\bibitem[\protect\citeauthoryear{{Nolan} et~al.,}{{Nolan} et~al.}{2012}]{2fgl}
{Nolan} P.~L.,  et~al., 2012, \mn@doi [The Astrophysical Journal Supplement
  Series] {10.1088/0067-0049/199/2/31}, 199, 31

\bibitem[\protect\citeauthoryear{{Ojha} et~al.,}{{Ojha} et~al.}{2010}]{Ojha10}
{Ojha} R.,  et~al., 2010, \mn@doi [\aap] {10.1051/0004-6361/200912724}, 519,
  A45

\bibitem[\protect\citeauthoryear{{Parsons} \& {Hinton}}{{Parsons} \&
  {Hinton}}{2014}]{2014APh....56...26P}
{Parsons} R.~D.,  {Hinton} J.~A.,  2014, \mn@doi [Astroparticle Physics]
  {10.1016/j.astropartphys.2014.03.002}, \href
  {https://ui.adsabs.harvard.edu/#abs/2014APh....56...26P} {56, 26}

\bibitem[\protect\citeauthoryear{{Piner} \& {Edwards}}{{Piner} \&
  {Edwards}}{2014}]{2014ApJ...797...25P}
{Piner} B.~G.,  {Edwards} P.~G.,  2014, \mn@doi [\apj]
  {10.1088/0004-637X/797/1/25}, \href
  {https://ui.adsabs.harvard.edu/abs/2014ApJ...797...25P} {797, 25}

\bibitem[\protect\citeauthoryear{{Piner} \& {Edwards}}{{Piner} \&
  {Edwards}}{2018}]{2018ApJ...853...68P}
{Piner} B.~G.,  {Edwards} P.~G.,  2018, \mn@doi [\apj]
  {10.3847/1538-4357/aaa425}, \href
  {https://ui.adsabs.harvard.edu/abs/2018ApJ...853...68P} {853, 68}

\bibitem[\protect\citeauthoryear{{Piranomonte}, {Perri}, {Giommi}, {Landt}  \&
  {Padovani}}{{Piranomonte} et~al.}{2007}]{2007A&A...470..787P}
{Piranomonte} S.,  {Perri} M.,  {Giommi} P.,  {Landt} H.,   {Padovani} P.,
  2007, \mn@doi [\aap] {10.1051/0004-6361:20077086}, \href
  {https://ui.adsabs.harvard.edu/#abs/2007A&A...470..787P} {470, 787}

\bibitem[\protect\citeauthoryear{{Piron} et~al.,}{{Piron} et~al.}{2001}]{Piron}
{Piron} F.,  et~al., 2001, \mn@doi [\aap] {10.1051/0004-6361:20010798}, \href
  {https://ui.adsabs.harvard.edu/#abs/2001A&A...374..895P} {374, 895}

\bibitem[\protect\citeauthoryear{{Pita} et~al.,}{{Pita}
  et~al.}{2014}]{2014A&A...565A..12P}
{Pita} S.,  et~al., 2014, \mn@doi [\aap] {10.1051/0004-6361/201323071}, \href
  {https://ui.adsabs.harvard.edu/#abs/2014A&A...565A..12P} {565, A12}

\bibitem[\protect\citeauthoryear{{Poole} et~al.,}{{Poole}
  et~al.}{2008}]{Poole2008}
{Poole} T.~S.,  et~al., 2008, \mn@doi [\mnras]
  {10.1111/j.1365-2966.2007.12563.x}, \href
  {http://adsabs.harvard.edu/abs/2008MNRAS.383..627P} {383, 627}

\bibitem[\protect\citeauthoryear{{Prokoph}, {Becherini}, {B{\"o}ttcher},
  {Boisson}, {Lenain}  \& {Sushch}}{{Prokoph}
  et~al.}{2015}]{2015ICRC...34..862P}
{Prokoph} H.,  {Becherini} Y.,  {B{\"o}ttcher} M.,  {Boisson} C.,  {Lenain}
  J.~P.,   {Sushch} I.,  2015, in 34th International Cosmic Ray Conference
  (ICRC2015). p.~862 (\mn@eprint {arXiv} {1509.03972})

\bibitem[\protect\citeauthoryear{{Roming} et~al.,}{{Roming}
  et~al.}{2005}]{Roming05}
{Roming} P.~W.~A.,  et~al., 2005, \mn@doi [\ssr] {10.1007/s11214-005-5095-4},
  \href {http://adsabs.harvard.edu/abs/2005SSRv..120...95R} {120, 95}

\bibitem[\protect\citeauthoryear{{Roming} et~al.,}{{Roming}
  et~al.}{2009}]{Roming09}
{Roming} P.~W.~A.,  et~al., 2009, \mn@doi [\apj] {10.1088/0004-637X/690/1/163},
  \href {http://adsabs.harvard.edu/abs/2009ApJ...690..163R} {690, 163}

\bibitem[\protect\citeauthoryear{{Sahu}, {L{\'o}pez Fort{\'\i}n}  \&
  {Nagataki}}{{Sahu} et~al.}{2019}]{2019ApJ...884L..17S}
{Sahu} S.,  {L{\'o}pez Fort{\'\i}n} C.~E.,   {Nagataki} S.,  2019, \mn@doi
  [\apjl] {10.3847/2041-8213/ab43c7}, \href
  {https://ui.adsabs.harvard.edu/abs/2019ApJ...884L..17S} {884, L17}

\bibitem[\protect\citeauthoryear{{Sanchez} \& {Deil}}{{Sanchez} \&
  {Deil}}{2015}]{enrico}
{Sanchez} D.,  {Deil} C.,  2015, {Enrico: Python package to simplify Fermi-LAT
  analysis}, Astrophysics Source Code Library (\mn@eprint {ascl} {1501.008})

\bibitem[\protect\citeauthoryear{{Shaw} et~al.,}{{Shaw}
  et~al.}{2013}]{2013ApJ...764..135S}
{Shaw} M.~S.,  et~al., 2013, \mn@doi [\apj] {10.1088/0004-637X/764/2/135},
  \href {http://adsabs.harvard.edu/abs/2013ApJ...764..135S} {764, 135}

\bibitem[\protect\citeauthoryear{{Skrutskie} et~al.,}{{Skrutskie}
  et~al.}{2006}]{Skrutskie06}
{Skrutskie} M.~F.,  et~al., 2006, \mn@doi [\aj] {10.1086/498708}, \href
  {https://ui.adsabs.harvard.edu/\#abs/2006AJ....131.1163S} {131, 1163}

\bibitem[\protect\citeauthoryear{{Stecker}, {de Jager}  \& {Salamon}}{{Stecker}
  et~al.}{1992}]{Stecker92}
{Stecker} F.~W.,  {de Jager} O.~C.,   {Salamon} M.~H.,  1992, \mn@doi [\apjl]
  {10.1086/186369}, \href {http://adsabs.harvard.edu/abs/1992ApJ...390L..49S}
  {390, L49}

\bibitem[\protect\citeauthoryear{{The Fermi-LAT collaboration}}{{The Fermi-LAT
  collaboration}}{2019}]{2019arXiv190210045T}
{The Fermi-LAT collaboration} 2019, arXiv e-prints, \href
  {https://ui.adsabs.harvard.edu/\#abs/2019arXiv190210045T} {p.
  arXiv:1902.10045}

\bibitem[\protect\citeauthoryear{{Thomas}, {Beuermann}, {Reinsch}, {Schwope},
  {Truemper}  \& {Voges}}{{Thomas} et~al.}{1998}]{1998A&A...335..467T}
{Thomas} H.~C.,  {Beuermann} K.,  {Reinsch} K.,  {Schwope} A.~D.,  {Truemper}
  J.,   {Voges} W.,  1998, \aap, \href
  {https://ui.adsabs.harvard.edu/#abs/1998A&A...335..467T} {335, 467}

\bibitem[\protect\citeauthoryear{{Voges} et~al.,}{{Voges}
  et~al.}{1999}]{1999yCat.9010....0V}
{Voges} W.,  et~al., 1999, VizieR Online Data Catalog, \href
  {https://ui.adsabs.harvard.edu/#abs/1999yCat.9010....0V} {p. IX/10A}

\bibitem[\protect\citeauthoryear{{Willingale}, {Starling}, {Beardmore},
  {Tanvir}  \& {O'Brien}}{{Willingale} et~al.}{2013}]{Willingale13}
{Willingale} R.,  {Starling} R.~L.~C.,  {Beardmore} A.~P.,  {Tanvir} N.~R.,
  {O'Brien} P.~T.,  2013, \mn@doi [\mnras] {10.1093/mnras/stt175}, \href
  {http://adsabs.harvard.edu/abs/2013MNRAS.431..394W} {431, 394}

\bibitem[\protect\citeauthoryear{{Wright} \& {Otrupcek}}{{Wright} \&
  {Otrupcek}}{1990}]{WO90}
{Wright} A.,  {Otrupcek} R.,  1990, in PKS Catalog (1990).

\bibitem[\protect\citeauthoryear{{Wright} et~al.,}{{Wright}
  et~al.}{2010}]{2010AJ....140.1868W}
{Wright} E.~L.,  et~al., 2010, \mn@doi [\aj] {10.1088/0004-6256/140/6/1868},
  \href {http://adsabs.harvard.edu/abs/2010AJ....140.1868W} {140, 1868}

\makeatother
\end{thebibliography}
